\documentclass[prd,preprintnumbers,nofootinbib,reprint,twocolumn]{revtex4-1}
\usepackage{amsmath,amssymb,bm,epsfig,color,graphicx}
\usepackage[mathscr]{eucal}
\usepackage{slashed}
\usepackage{cancel}
\usepackage{yhmath}
\usepackage[colorlinks=true,linkcolor=blue,citecolor=blue,urlcolor=blue]{hyperref}
\usepackage[caption=false]{subfig}
\usepackage{hyperref}
\usepackage{tikz,xcolor}

\usepackage{amsmath}
\usepackage{slashed}
\usepackage{comment}
\usepackage{cancel}
\usepackage{multirow}
\usepackage{tikz-feynman}
\tikzfeynmanset{compat=1.1.0}

\newcommand{\bea}{\begin{array}}
\newcommand{\eea}{\end{array}}
\newcommand{\beq}{\begin{eqnarray}}
\newcommand{\eeq}{\end{eqnarray}}

\newcommand{\del}{\partial}

\def\thf{\theta_f}
\def\vpf{\varphi_f}
\def\thfb{\theta_{\bar{f}}}
\def\vpfb{\Delta \varphi}

\newcommand{\indot}[2]{#1\cdot#2}

\definecolor{orange}{RGB}{255,100,0}
\definecolor{rosepink}{RGB}{248,100,100}

\begin{document}

\title{Light Dilaton in Rare Meson Decays and  Extraction of its CP Property  \\
}
\preprint{KEK--TH--2569}

\author{Sudhakantha Girmohanta$^{1,2}$, Yuichiro Nakai$^{1,2}$, Yoshihiro Shigekami$^{1,2}$ and Kohsaku Tobioka$^{3,4}$
\\*[10pt]
$^1${\it \normalsize Tsung-Dao Lee Institute, Shanghai Jiao Tong University, \\
520 Shengrong Road, Shanghai 201210, China} \\*[3pt]
$^2${\it \normalsize School of Physics and Astronomy, Shanghai Jiao Tong University, \\
800 Dongchuan Road, Shanghai 200240, China} \\*[3pt]
$^3${\it \normalsize Department of Physics, Florida State University, Tallahassee, Florida 32306, USA} \\*[3pt]
$^4${\it \normalsize Theory Center, High Energy Accelerator Research Organization (KEK), Tsukuba 305-0801, Japan} \\*[5pt]
}

\begin{abstract}

The dilaton $\phi$ is a pseudo-Nambu-Goldstone boson associated with the spontaneous breaking of scale invariance in a nearly conformal theory, and couples to the trace of the stress-energy tensor. 
We analyze experimental constraints on a light dilaton with mass in the MeV-GeV range from rare meson decays. 
New model-independent inclusive bounds for the $b \to s \phi$ transition largely exclude the parameter space of a light dilaton that could explain the muon $g-2$ anomaly. 
Despite similarities between a dilaton and a Higgs-portal scalar, the dilaton-photon coupling is enhanced compared to the Higgs-portal scalar due to contributions from loops of the conformal sector. 
Consequently, the shortened lifetime of the dilaton relaxes bounds from $K \to \pi$ + invisible searches at the NA62 experiment and constraints from the Big Bang Nucleosynthesis. 
We utilize this fact to search for the dilaton signature at a lepton collider such as the ongoing Belle II experiment. 
Further, we demonstrate how to extract the CP property of the dilaton using the variation of the differential cross-section of $e^+ e^- \to e^+ e^- \phi$ with the azimuthal angle between the outgoing leptons. 

\end{abstract}

\maketitle

\section{Introduction}
\label{sec:intro}

The existence of approximate scale invariance with dynamical spontaneous breaking is an intriguing possibility of physics beyond the Standard Model (SM) that is able to address the quantum instability and hierarchy of the electroweak scale with respect to the Planck scale. 
The AdS/CFT correspondence~\cite{Maldacena:1997re,Gubser:1998bc,Witten:1998qj} (see also refs.~\cite{Arkani-Hamed:2000ijo,Rattazzi:2000hs}) tells us that such a 4D nearly-conformal field theory is holographically dual to the 5D Randall-Sundrum (RS) model with a warped extra dimension bounded by two 3-branes, called UV and IR branes~\cite{Randall:1999ee}. 
Physics on the IR brane is redshifted due to an exponential warp factor. 
Here, the existence of the IR brane in the 5D theory corresponds to the spontaneous breaking of scale invariance (SBSI). 
The RS model has a radion degree of freedom to describe the distance between the two 3-branes. 
This radion plays the role of a pseudo-Nambu-Goldstone boson (pNGB) associated with SBSI, called $dilaton$, in the nearly-conformal theory. 
The radion (dilaton) mass depends on a mechanism to stabilize the brane distance, while typically it is somewhat smaller than the mass scale of the IR brane (the scale of SBSI), which is set to the TeV scale. 
However, as pointed out in refs.~\cite{Bellazzini:2013fga,Coradeschi:2013gda}, a lighter dilaton can be achieved, and it has a rich phenomenology~\cite{Abu-Ajamieh:2017khi} as the dilaton couples to SM particles through the trace of the stress-energy tensor. 

The radion/dilaton emerges as the lightest degree of freedom from warped extra-dimensional models, or by duality, from composite dynamics breaking a nearly conformal theory. 
Thus, it serves as the key probe into these class of theories for a low-energy observer. 
In spite of the efforts in elucidating dilaton phenomenology at high energy colliders~\cite{Csaki:2000zn,Csaki:2007ns}, a detailed phenomenological study was lacking for the MeV-GeV mass range. 
Ref.~\cite{Abu-Ajamieh:2017khi} partly filled this gap, motivated by the fact that a dilaton in this mass range can address the muon $g-2$ anomaly~\cite{Chen:2015vqy,Marciano:2016yhf,Muong-2:2006rrc,Keshavarzi:2018mgv,Muong-2:2021ojo, Muong-2:2023cdq}. 
Borrowing the axion search constraints, it was claimed that a direct coupling of the radion to fermions weakens many bounds and allows a region of the parameter space that explains the muon $g-2$ anomaly. 
Nevertheless, a dedicated detailed study for the dilaton phenomenology is missing for this mass range. 
In particular, we note that a dilaton coupling to leptons implies a dilaton coupling to weak gauge bosons. 
Specifically, there is a tree-level interaction between the dilaton and the $W$ boson as the $W$ boson mass breaks the scale invariance. 
Besides, one would also expect dilaton couplings to quarks. 
Consequently, it is unclear whether the parameter space for the muon $g-2$ anomaly is viable or not when confronted with stringent rare meson decay constraints{\footnote{The Kaluza-Klein (KK) mode contribution is typically too small to address the muon $g-2$ anomaly~\cite{Beneke:2014sta}.}} (see ref.~\cite{Goudzovski:2022vbt} for a recent review on rare meson decay phenomenology). 
Due to reasons we will illustrate later, experimental constraints on a dilaton resemble those on a Higgs-portal scalar~\cite{Winkler:2018qyg}. 
However, there is a key difference, namely that the dilaton has an enhanced coupling to the photon due to loops of the conformal sector~\cite{Csaki:2007ns}. 
We will thoroughly examine the constraints for the radion/dilaton in the MeV-GeV mass range in detail, which will clarify the current status.

In the first part of the present paper, we will investigate $B \rightarrow K\phi$ and $K \to \pi \phi$ transitions for a light dilaton $\phi$ with mass in the MeV-GeV range. 
We will calculate the dilaton production from the $B \rightarrow K\phi$ process at the one-loop level when a dilaton directly couples to quarks as well as weak gauge bosons. 
It will be shown that the current experimental data for the inclusive $b \rightarrow s \phi$ transition largely exclude the dilaton parameter space to explain the muon $g-2$ anomaly. 
Further, in part of the light dilaton parameter space, constraints from searches for $K \to \pi +$ invisible modes in the NA62 experiment at CERN are relaxed due to its enhanced coupling with the photon. 
The constraint from the Big Bang Nucleosynthesis (BBN) is also comparably weaker due to the shortened lifetime of the dilaton. 
On the other hand, searches for the visible signature modes of dilaton decays such as $K_{L} \to \pi^0 \gamma \gamma$ and $K_{L} \to \pi^0 e^+ e^-$ at the KTeV experiment constrain the dilaton-photon coupling from a complementary direction in the parameter space. 
With mild assumption on the relaxed direct coupling to electrons, a window for the light dilaton appears that is consistent with the current stringent rare meson decay constraint and has a large enough coupling to photons that can be searched for in modes like $e^+ e^- \to \ell^+ \ell^- \phi$ where $\ell = e, \mu$ at the Belle II experiment. 

The results of our analysis for the dilaton parameter space and its heightened coupling to the photon provide the motivation to proceed one step further and extract the CP property of the dilaton in a prospective future signal and analyze its distinction thereof from axions. 
In the second part of the paper, we pursue this direction. 
An imprint of the CP nature of $\phi$ remains on the distribution of the differential cross-section with a Lorentz-invariant angle $\Delta \varphi$ to be defined later. 
We utilize a kinematical method to illustrate the distinct signature of a CP-odd and CP-even particle, while thanks to the enhanced dilaton-photon coupling, enough statistics can be obtained, which is not possible for a Higgs-portal scalar in the focused light mass range. 
Our method is not particular to Belle II and can be taken over by any lepton collider. 

The organization of the rest of the paper is given as follows. 
In section~\ref{sec:model}, we define the model and summarize the dilaton mass and couplings to the SM. 
Then, section~\ref{sec:constraints} discusses the constraints from rare meson decays and collider searches for a light dilaton with mass in the MeV-GeV range. 
Their implications for the dilaton interpretation of the muon $g-2$ anomaly are discussed. 
In section~\ref{sec:obs}, we explore the extraction of the CP nature. 
Section~\ref{sec:sum} is devoted to conclusions and discussions, while appendix~\ref{sec:calc} contains some relevant details of the kinematical calculations.

\section{The Model}\label{sec:model}

The low-energy effective description of a 4D nearly-conformal field theory is expressed in terms of the dilaton $\phi$, which is the CP-even pNGB of the broken scale invariance. 
Under a scale transformation $x^{\mu} \to x'^{\mu} =  e^{-\omega} x^\mu$, below the conformal breaking scale $f_\phi$, the dilaton undergoes a shift transformation as $\phi (x) \to \phi'(x') = \phi(x)+ \omega f_{\phi}$. 
Here, $\phi$ can be conveniently embedded in a conformal compensator $\chi(x) \equiv f_\phi e^{\phi(x)/f_\phi}$, such that under the scale transformation the compensator transforms linearly $\chi(x) \to \chi'(x') = e^{\omega} \chi(x)$. 
Then, the dilaton effective Lagrangian can be constructed so that it is symmetric under the above realization of the scale transformation. 
As the breaking of the dilatation current is given by the trace of the energy-momentum tensor $T^{\mu \nu}$, the coupling of the $\chi$ would be such that it compensates for the breaking, corrected by explicit conformal breaking effects. 
Therefore, the action of the dilaton can be written as~\cite{Chacko:2012sy, Abu-Ajamieh:2017khi}
    \begin{equation}
        S_{\phi, {\rm eff}} = \int d^4 x \left[\frac{1}{2}(\partial_\mu \phi)(\partial^{\mu} \phi) -V_{\rm eff} (\phi) + c \frac{\phi}{f_\phi} T^{\mu}_{\mu} \right] \ ,
        \label{phi_action:eq}
    \end{equation}
where we assume the kinetic term of $\phi$ is canonically normalized, $V_{\rm eff}$ contains a potential generated for $\phi$ from explicit violation of the conformal invariance, and $c$ denotes some constant that depends on how the scale invariance is broken by $T^{\mu}_\mu$.

\subsection{Dilaton interactions}

Let us now summarize dilaton interactions with the SM particles. 
The coupling to a massive vector boson $V^\mu$ can be obtained by noting that the scale invariance is broken by the mass of the vector boson $m_V$. 
Then, the dilaton effective coupling in the unitary gauge must be 
\begin{equation}
    {\cal L}_{{\rm eff}, \chi V} = \left(\frac{\chi}{f_\phi}\right)^2 m_V^2 V_\mu V^\mu \ .
    \label{Vchi:eq}
\end{equation}
Expanding $\chi = f_\phi e^{\phi/f_\phi}$ in terms of $\phi$, one obtains, to the leading order,
\begin{equation}
    {\cal L}_{{\rm eff}, \phi V} = 2 \left( \frac{\phi}{f_\phi} + \frac{\phi^2}{f_\phi^2} \right) m_V^2 V_\mu V^\mu \ .
    \label{Vphi:eq}
\end{equation}
Similarly, the scale invariance is broken by the mass of a fermion $f$, and hence, the dilaton couples to the fermion as
\begin{equation}
    {\cal L}_{{\rm eff}, \chi f} = -\left(\frac{\chi}{f_\phi}\right) m_f \bar f f \ ,
\end{equation}
with the fermion mass $m_f$. 
To the linear order, we obtain
\begin{equation}
    {\cal L}_{{\rm eff}, \phi f} = -c_{\phi f f}\frac{\phi}{f_\phi} m_f \bar f f \ ,
    \label{eq:dilaton_fermion_coupling}
\end{equation}
where we have introduced $c_{\phi f f}$, a dimensionless parameter that depends on the mixing between elementary and composite sectors in the UV theory, which is related to the anomalous dimension of the fermion and used to obtain the flavor structure of the SM. 
In the dual 5D picture, it amounts to the specification of the wavefunction profile of the fermion in the warped extra dimension. 
We are interested in a general analysis, and simply parameterize the model dependence in term of the coefficient $c_{\phi f f}$. 
The explicit violation of the conformal invariance by a nearly marginal operator generating a dilaton mass $m_{\phi}$ corrects the above interactions by ${\cal O}(m_\phi^2/f_\phi^2)$, which we neglect in the following discussion. 

Although a massless gauge field possesses the traceless energy-momentum tensor and no scale dependence at the classical level, the running of the gauge coupling generates a scale dependence. 
Such a running receives contributions from elementary and strongly interacting states above the breaking scale of the scale invariance, and elementary and other light emergent states from the strong sector below the breaking scale. 
With the $\beta$-function coefficient above and below the symmetry breaking scale denoted as $b_{+}$ and $b_{-}$ respectively, the coupling of the dilaton to a massless gauge boson can be written as~\cite{Chacko:2012sy}
\begin{equation}
    {\cal L}_{{\rm eff}, \phi A} = \frac{g^2}{32 \pi^2} (b_{-}-b_{+}) \frac{\phi}{f_{\phi}} F^{\mu \nu} F_{\mu \nu} \ ,
    \label{LphiA:eq}
\end{equation}
where $F$ is the field strength tensor of the massless gauge field $A^\mu$, and $g$ denotes the corresponding gauge coupling. Explicitly, the CFT degrees of freedom contribute to the 2-point function of the massless fundamental gauge field~\cite{Csaki:2007ns}, such that 
\begin{equation}
    b_{+} = -\frac{8 \pi^2}{g^2 \ln {\Lambda_{\rm UV}/f_\phi}} \ ,
\end{equation}
where $\Lambda_{\rm UV}$ is the UV cut-off of the theory. Therefore, the dilaton coupling to a massless gauge boson can be summed up as follows
\begin{equation}
    {\cal L}_{{\rm eff}, \phi A} = \left[ \frac{1}{4 \ln {\Lambda_{\rm UV}/f_\phi}} +  \frac{g^2}{32 \pi^2} b_{-} \right] \frac{\phi}{f_\phi} F^{\mu \nu} F_{\mu \nu} \ .
    \label{eq:dilaton_photon_coupling}
\end{equation}
For concreteness, we will take the cut-off at an intermediate scale, namely $\Lambda_{\rm UV} \sim 10^{10} \  {\rm GeV}$, while it is straightforward to do an analysis with other choices.

Finally, the mixing between the SM Higgs field and the dilaton can only arise from explicit conformal breaking effects
and hence is quite suppressed. This is a non-trivial consequence of the scale invariance
and understood as follows. The Higgs-dilaton potential has a restrictive form due to the scale invariance, namely
\begin{equation}
    V(\chi,h) = \left( \frac{\chi}{f_\phi}\right)^4 V_0 (h) \ ,
    \label{Higgs_dilaton_pot:eq}
\end{equation}
where $h$ is the SM Higgs field, and $V_0$ only contains terms involving $h$.
Therefore, upon expanding $V_0$ around the minimum,
no linear term in $h$, that could be responsible for a Higgs-dilaton mixing, appears.
Only such effect can arise from explicit conformal breaking terms, and hence, the corresponding mixing angle will be suppressed by~\cite{Chacko:2012sy}
\begin{equation}
    \theta_{\phi h} \sim \left( \frac{m_\phi^2}{16 \pi^2 f_\phi^2} \right) \frac{v}{f_\phi} \ll 1\ ,
\end{equation}
where $v$ is the Higgs vacuum expectation value (VEV). 

Finally, let us compare and contrast the dilaton interactions with those of a Higgs-portal scalar. A common way to extend the SM with a real singlet scalar ($S$) is to assume that it mixes with the SM Higgs doublet $H$ through the following term~\cite{Winkler:2018qyg, Kachanovich:2020yhi}
\begin{equation}
{\cal L}_{Sh} = \left( g_{1S} S + g_{2S} S^2 \right) \left( H^\dagger H \right) \ .
\label{eq:LSh}
\end{equation}
After mass diagonalization, $S$ inherits the SM Higgs interactions suppressed by the sine of the Higgs-scalar mixing angle ($\theta_{Sh}$). For example, the singlet coupling to a SM fermion $f$ is
\begin{equation}
{\cal L}_{Sf} = - \left( \frac{\sin \theta_{Sh}}{v} \right) m_f \bar f f S \ .
\label{eq:LSf}
\end{equation}
Although the dilaton-fermion coupling arises from a different origin, namely SBSI instead of mixing with the physical Higgs, the effective Lagrangian~\eqref{eq:LSf} looks similar to Eq.~\eqref{eq:dilaton_fermion_coupling}. However, there is an important difference in the factor $c_{\phi f f}$ that depends on the anomalous dimension of $f$. Another key difference is in the coupling with the photon. While $S$ obtains the photon coupling through loops of SM particles, especially $t, W$~\cite{Marciano:2011gm}, the dilaton also gets additional contribution from loops of the conformal sector as shown in Eq.~\eqref{eq:dilaton_photon_coupling}. These facts will play crucial roles in our analysis of the dilaton phenomenology.

\subsection{Dilaton mass}

In this section, we will illustrate how the dilaton receives its mass, and outline how a GeV-scale dilaton can emerge from the underlying theory.
The dilaton receives its mass as a result of an explicit violation of the conformal symmetry.
A general analysis of the dilaton mass can be carried out by assuming that the initial conformal theory is perturbed
by a nearly marginal operator of mass dimension $4-\epsilon$,
\begin{equation}
    {\cal L}_{\rm CFT} \to {\cal L}_{\rm CFT} + c_{{\cal O}} {\cal O} \ .
\end{equation}
The scaling of this operator depends on the coupling $c_{{\cal O}}$ and
how it evolves with the renormalization group (RG) flow.
At the breaking scale of the conformal theory, the scaling dimension of the operator will dictate the dilaton mass. 

Alternatively, we can consider a dual 5D picture of the RS model.
Here, the ${\mathbb R}_4 \otimes S_1/{\mathbb Z}_2$ metric with the graviton $g_{\mu \nu}$ and the modulus field $T(x)$
to describe the distance between two 3-branes is 
\begin{equation}
    ds^2 = G_{AB} dx^A dx^B = e^{-2 k |\theta| T(x)} g_{\mu \nu} (x) dx^\mu dx^\nu - T(x)^2 d\theta^2 \ , 
\end{equation}
where $x^\mu$ denotes the 4D coordinate, $k$ is the curvature scale,
and $\theta \in [-\pi, \pi]$ parameterizes the extra dimension
with the UV (IR) brane placed at the fixed point $\theta=0$ ($\theta=\pi$).
Integrating out the extra dimension, the 4D effective action is given by
\begin{align}
    \nonumber
    S_{\phi, \rm eff} \supset& \, \frac{2 M_*^3}{k} \int d^4 x \sqrt{-g}  \left( 1 -\left({\phi}/{Z}\right)^2 \right) R_4 \\ & + \frac{1}{2} \int d^4 x \sqrt{-g} \, \partial_\mu \phi(x) \partial^\mu \phi(x)  \ ,
    \label{radion_action:eq}
\end{align}
where $M_*$ is the 5D Planck constant, $R_4$ is the Ricci scalar constructed from
$g_{\mu \nu}$, and $\phi(x) \equiv Z e^{-\pi k T(x)}$ with $Z\equiv\sqrt{24 M_*^3/k}$
denotes the canonically normalized radion. Note that $N \equiv 4 \pi (M_*/k)^{3/2}$ is the order of number of colors in the dual large-$N$ CFT.
In Eq.~\eqref{radion_action:eq}, the radion is massless. 

To generate a radion mass, we can utilize the classic Goldberger-Wise mechanism~\cite{Goldberger:1999uk}
where a 5D scalar field $\Phi$ with bulk mass $m_{\Phi}$ is introduced.\footnote{
For another mechanism of radion stabilization, see ref.~\cite{Fujikura:2019oyi}.
}
It feels brane localized potentials,
\begin{align}
\label{GWscalar}
    \nonumber
    S_{\Phi} =& \int d^4 x d\theta \bigg[ \frac{\sqrt{G}}{2}\left( G^{MN} \partial_M \Phi \partial_N \Phi
     -m_{\Phi}^2 \Phi^2\right) \\
     & -\sqrt{-g_h}\lambda_h (\Phi^2 - v_h^2)^2 - \sqrt{-g_v}\lambda_v (\Phi^2 - v_v^2)^2 \bigg]  \ .
\end{align}
Here, $g_{h}$ ($g_v$) is the induced metric at the UV (IR) brane, and the mass dimensions for $\Phi, v_v, v_h$ are $3/2$, whereas $\lambda_{h,v}$ carry mass dimension $-2$. When $\lambda_{v,h}$ are sufficiently large,
it is energetically favorable for $\Phi$ to have a fixed VEV $v_h$ ($v_v$) at the UV (IR) brane.
The solution for the classical equation of motion for $\Phi$ is then given by
\begin{equation}
    \Phi (\theta) = A e^{ (4+\epsilon) k T |\theta|} + B e^{-\epsilon k T |\theta|} \ , 
    \label{Phi_sol:eq}
\end{equation}
where $\epsilon \equiv \sqrt{4+m_{\Phi}^2/k^2}-2 \ll 1$ is the parameter that controls the explicit breaking of the conformal invariance due to the bulk scalar mass, and the coefficients $A, B$ are determined from the VEVs for $\Phi$ at the branes. 
Plugging Eq.~\eqref{Phi_sol:eq} back into the action~\eqref{GWscalar} and integrating out the extra dimension,
one gets the effective radion potential, namely
\begin{equation}
    V_{\rm eff}(\phi) \simeq  k \left( \frac{\phi}{Z} \right)^4 v_v^2 \Big[ (4+2 \epsilon) \bigg\{1 - \frac{v_h}{v_v} \left(\frac{\phi}{Z} \right)^\epsilon\bigg\}^2-\epsilon \Big] \ .
\end{equation}
The potential has a global minimum at 
\begin{equation}
    \frac{\langle \phi \rangle}{Z} =\bigg[ \frac{v_h}{v_v} X_{\rm min} \bigg]^{1/\epsilon} \ ,
    \label{phi_vev:eq}
\end{equation}
where 
\begin{align}
\nonumber
    X_{\rm min} &= \frac{1}{4+2 \epsilon} \Big[ 4+ {\epsilon} + \sqrt{\epsilon \left(4+ \epsilon \right)} \Big] \\
  &=  1 + \frac{\sqrt{\epsilon}}{2} - \frac{\epsilon}{4} + {\cal O}(\epsilon^{3/2}) \ .
\end{align}
Therefore, the radion mass is given by
\begin{align}
    m_{\phi}^2  &= \frac{\partial^2 V_{\rm eff}(\phi)}{\partial\phi^2} \bigg|_{\langle\phi\rangle} \nonumber \\
    &= \frac{2 v_v^2}{3 M_*^3} M_{\rm IR}^2 \epsilon^{3/2} \left(1+\frac{\sqrt{\epsilon}}{2}\right)
    +{\cal O}(\epsilon^{5/2}) \ .
    \label{radion_mass_naive:eq}
\end{align}
Here, we have defined the IR mass scale $M_{\rm IR} \equiv k e^{- \pi k \langle T \rangle}$
with the modulus VEV $\langle T \rangle$ obtained from Eq.~\eqref{phi_vev:eq}. Here, $f_\phi \simeq \sqrt{6} M_{\rm IR}$~\cite{Csaki:2000zn}.

For a more rigorous treatment, ref.~\cite{Csaki:2000zn} solved the coupled equations of the metric fluctuations
and bulk scalar stabilizing the system for a general class of potentials,
and concluded that the radion mass has a similar form as the naive approximation~\eqref{radion_mass_naive:eq} except that it scales as $\epsilon^2$. 
Then, the radion mass is suppressed by the small parameter $\epsilon v_v/M_*^{3/2}$ compared to the IR mass scale.
Concretely, for $k \simeq M_{\rm Pl}$, with the correct scaling for $\epsilon$, one obtains
\begin{equation}
    m_{\phi} \simeq 1 \  {\rm GeV} \left( \frac{\epsilon}{1/40}\right) \left(\frac{v_v/M_*^{3/2}}{0.07} \right) \left( \frac{M_{\rm IR}}{1 \  {\rm TeV}} \right) \ ,
\end{equation}
where to keep $M_{\rm IR}$ fixed, $v_h/v_v$ is chosen to be $2.4$ {\footnote{See refs.~\cite{Coradeschi:2013gda, Bellazzini:2013fga, Abu-Ajamieh:2017khi} for mechanisms that can generate an even smaller dilaton mass naturally.}}.

\subsection{Dilaton contribution to muon $g-2$}
\label{sec:muong-2}

The anomalous magnetic moment of the muon ($a_\mu \equiv {(g_\mu -2)}/{2}$) has been measured precisely by BNL E821 experiment~\cite{Muong-2:2006rrc}, and improved by Fermilab~\cite{Muong-2:2023cdq} with a precision of $0.20$ parts per million. Therefore, it serves as a precision probe beyond the SM. Comparison of the SM theory inferred value is in tension with the experimental value obtained, although precise determination of SM hadronic vacuum polarization contribution from lattice calculation remains an important pursuit. Evidently, as the dilaton has a coupling to the muon, the one-loop diagram has a contribution to the anomalous magnetic moment of the muon $a_\mu$. 
To the leading order (LO), 
\begin{align}
\Delta a_{\mu} =
{\displaystyle  \frac{c_{\phi \mu \mu}^2 m_{\mu}^2}{8 \pi^2 f_{\phi}^2} r_{\phi}^2 \int_0^1 \! dx \frac{x^2 \left( 2 - x \right)}{r_{\phi} x^2 - x + 1}} \, ,
\end{align}
where  $r_{\phi} \equiv m_{\mu}^2 / m_{\phi}^2$. 
The current discrepancy of $\Delta a_{\mu} \equiv a_\mu^{\rm exp}-a_\mu^{\rm SM}$ is~\cite{Muong-2:2006rrc,Keshavarzi:2018mgv,Aoyama:2020ynm,Muong-2:2021ojo,Muong-2:2023cdq}
\begin{align}
\Delta a_{\mu} = \left( 2.49 \pm 0.48 \right) \times 10^{-9} \ .
\end{align}
Here we have used the theoretical calculation of $a_\mu$ from the 2020 white paper~\cite{Aoyama:2020ynm}, and for the experimental value, we take the world average including the recently released FNAL Run-2 and Run-3 data~\cite{Muong-2:2023cdq}. Note that the additional contribution due to a light dilaton, thanks to its CP-property, has the desirable sign, whereas the LO contribution of the axion has the opposite sign (see ref.~\cite{Marciano:2016yhf} for NLO contributions). Motivated by this, it has been argued that a light radion coupling to muons can alleviate the muon $g-2$ anomaly~\cite{Abu-Ajamieh:2017khi}. However, this has to be confronted with stringent laboratory constraints, especially coming from rare meson decays, which will be elucidated in the next section.

\section{The dilaton parameter space}
\label{sec:constraints}

In this section, we consider the experimental and astrophysical constraints on the parameter space of the dilaton. Most stringent constraints come from searches for decays of $K$ and $B$ mesons, while for a larger dilaton mass collider constraints become relevant. We will comment on bounds from BBN and supernovae 1987a.  As alluded to earlier, although the dilaton has a similar coupling to the SM as a Higgs-portal scalar, there are a few important differences that affect its phenomenology which we elaborate on here. Our results are shown in Fig.~\ref{fig:dilaton_plot}. 

\begin{figure}[t!]
    \centering
    \includegraphics[width=0.9\linewidth]{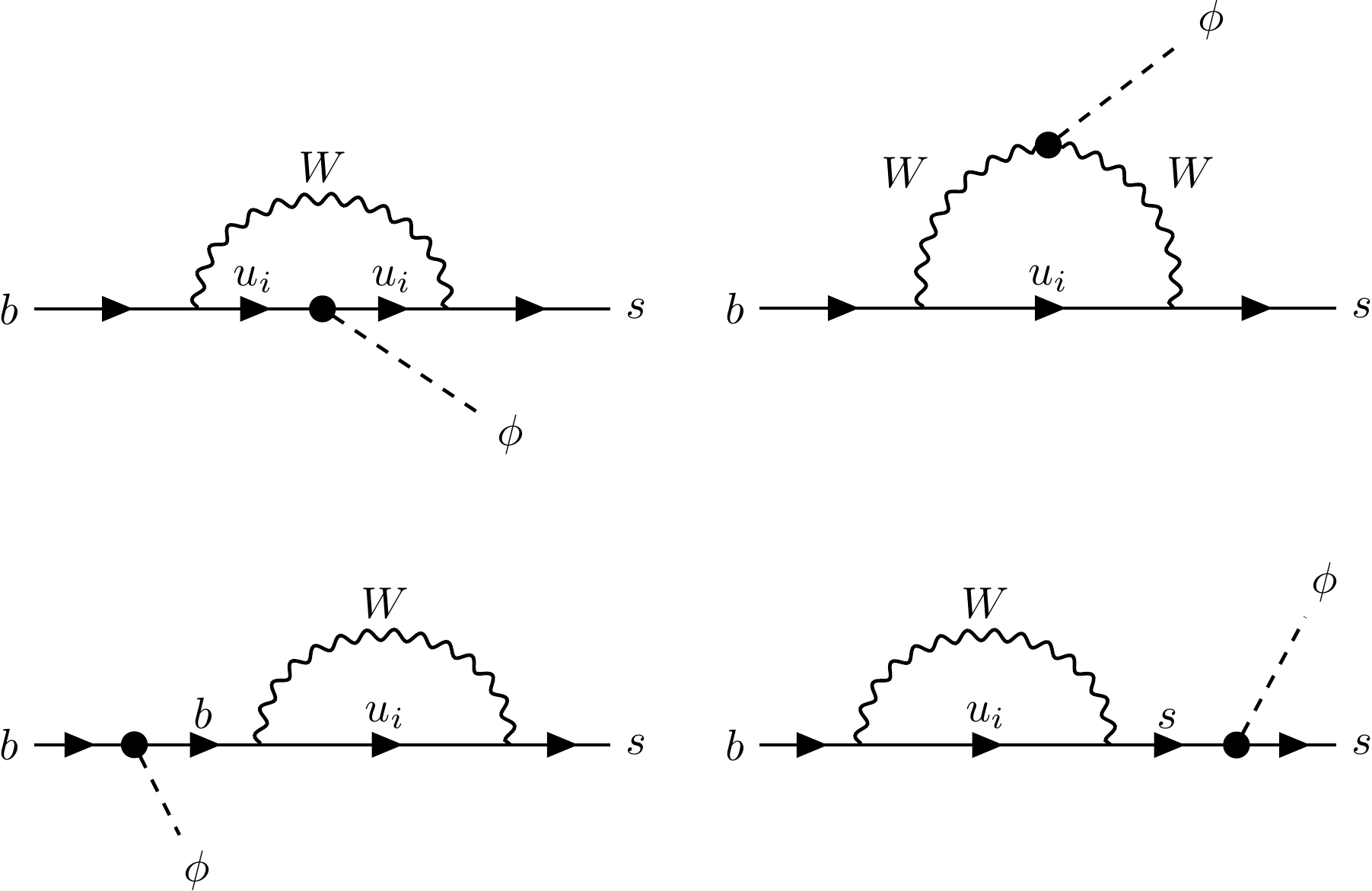}
    \caption{1PI diagrams mediating $b \to s \phi$ in the unitary gauge.}
    \label{fig:btosphi}
\end{figure}

\subsection{Constraints from rare meson decays}
\label{sec:Binclusive}
\begin{figure*}[ht!]
    \centering
    \includegraphics[width=0.85\textwidth]{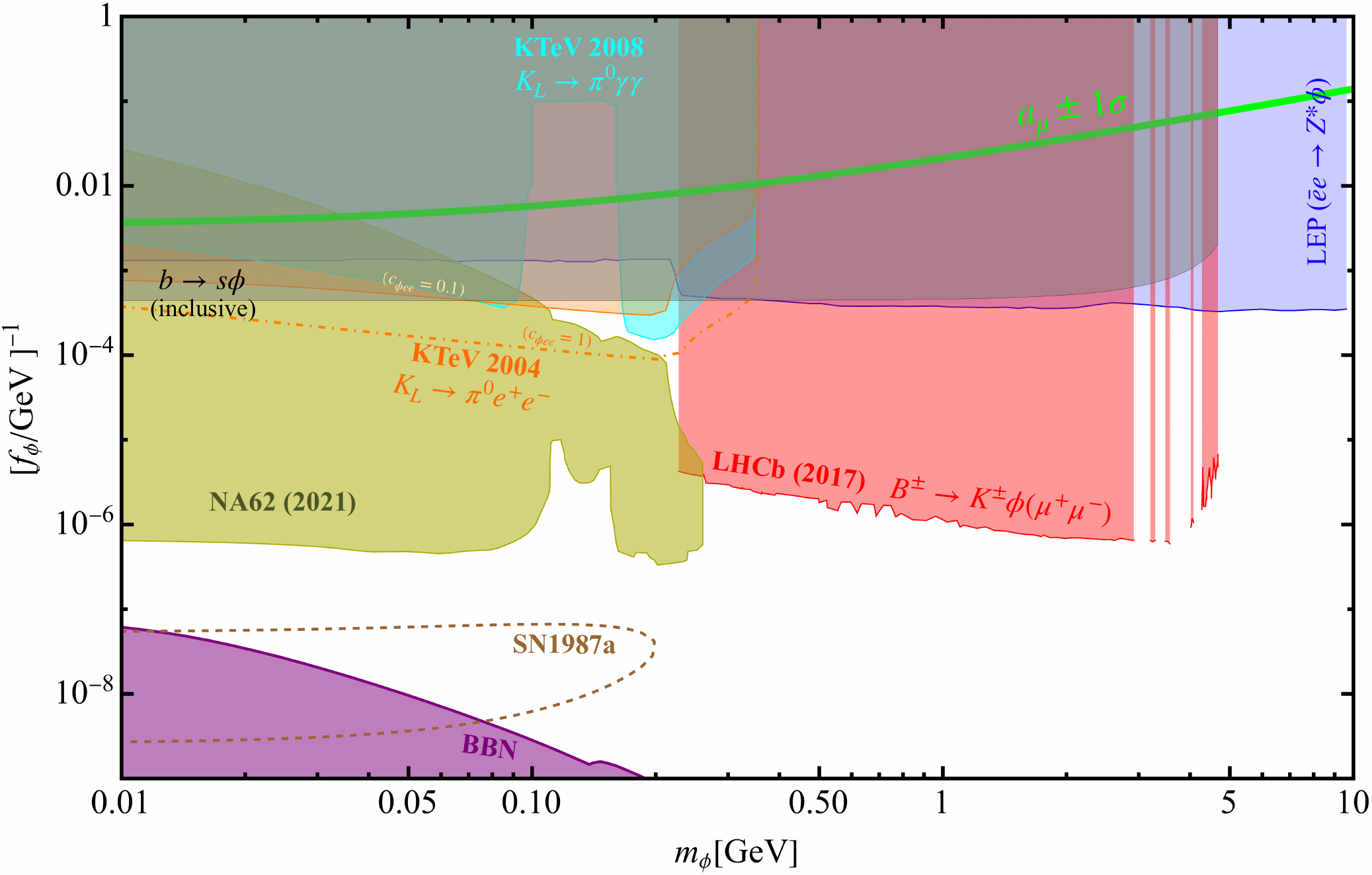}
    \caption{Constraints from rare meson decays, collider, cosmology, and astrophysics on the parameter space of a light dilaton. The cyan region represents the KTeV bound on $K_{\rm L} \to \pi^0 \phi (\to \gamma \gamma)$~\cite{KTeV:2008nqz}, while $K_{\rm L} \to \pi^0 \phi (\to e^+ e^-)$~\cite{KTeV:2003sls} constraint is depicted in orange for $c_{\phi e e}=0.1, 1$ respectively. The darker yellow region corresponds to NA62 bound on the mode $K^+ \to \pi^+ + {\rm inv}$~\cite{NA62:2021zjw} (where $\phi$ escapes the detector with an effective invisible signature). The dilaton-photon coupling is calculated from Eq.~\eqref{eq:dilaton_photon_coupling} with $\Lambda_{\rm UV}=10^{10}$ GeV. The black-shaded region denotes inclusive bounds from $b \to s \phi$ decay. The purple and brown dashed regions stand for bounds from BBN and supernova SN1987a respectively. The red region shows the constraint from semileptonic $B \to K  \phi(\to \mu^+ \mu^-)$ decay from LHCb~\cite{LHCb:2016awg}, while the blue region corresponds to exclusion from LEP searches for $e^+ e^- \to Z^* \phi$~\cite{L3:1996ome}. Smaller $c_{\phi \mu \mu}$ can reduce $\phi \to \mu^+ \mu^-$ branching ratio as it does not affect hadronic decay channels, easing the semileptonic $B$ meson decay constraint. In the green band, the dilaton contribution to the muon $g-2$ can bring $\Delta a_\mu$ within $1\sigma$ deviation from the experiment. Note that the enhanced coupling to photons for the light dilaton plays a crucial role in opening up a window of parameters below the muon threshold region.}
    \label{fig:dilaton_plot}
\end{figure*}

If the mass of $\phi$ is less than the threshold where the $B \to K \phi$ decay with on-shell $\phi$ is allowed, we need to take into account constraints from the $B$ decay processes. 
In Fig.~\ref{fig:btosphi}, we show the relevant one-loop diagrams to the $b \to s$ transition with the dilaton $\phi$. 
These contributions can be summarized in the following effective couplings:
\begin{align}
\nonumber
\mathcal{L}_{{\rm eff}, \phi s b} \supset & \frac{4 G_F}{\sqrt{2}} V_{u_i b} V_{u_i s}^* \Bigl( g_R^B (q^2) \bar{s}_L b_R \phi + g_L^B (q^2) \bar{s}_R b_L \phi \Bigr)\,  \\ & + {\rm h.c.} , \label{eq:btosphi}
\end{align}
where $G_F$ is the Fermi constant, $V_{i j}$ is an element of the Cabibbo-Kobayashi-Maskawa (CKM) matrix, $q$ is the four-momentum of $\phi$, and hence, $q^2 = m_{\phi}^2$ for the on-shell $\phi$. 
Note that due to the proper chirality flipping on the fermion line, $g_R^B (q^2)$ is expected to be larger than $g_L^B (q^2)$ by a factor of $m_b / m_s \approx 50$. 
Therefore, hereafter, we focus on the coupling $g_R^B (q^2)$ only. 

The calculation for the $g_{R, L}^B (q^2)$ is well-known~\cite{Batell:2009jf,Winkler:2018qyg} (see \cite{Kachanovich:2020yhi} for a general $R_\xi$ gauge calculation).  
With $c_{\phi q q } = 1$, and on-shell $\phi$, one obtains
\begin{align}
g_R^B = \frac{3 m_b m_t^2}{32 \pi^2 f_{\phi}} \, .
\end{align} 
Using this coupling, we can estimate the branching ratio for $B \to K \phi$. 
For this purpose, we need to calculate the hadronic matrix elements $\langle K | \bar{s} b | B \rangle$, $\langle K | \bar{s} \gamma^5 b | B \rangle$. 
From ref.~\cite{Ball:2004ye}, we get
\begin{align}
\langle K | \bar{s} b | B \rangle &= \frac{M_B^2 - M_K^2}{m_b - m_s} f_0^K (q^2) \, , \\[0.5ex]
\langle K | \bar{s} \gamma^5 b | B \rangle &= 0 \, ,
\end{align}
where $f_0^K(q^2) = \frac{0.33}{1-q^2/(37.5 \ {\rm GeV}^2)}$ is the form factor for the $B \to K$ transition~\cite{Ball:2004ye}. 
Therefore, we have the following branching ratio for $B \to K \phi$ process:
\begin{align}
\nonumber
{\rm BR} (B \to K \phi) = &\frac{\left| g_R^B \right|^2}{64 \pi \Gamma_B} \left( \frac{4 G_F}{\sqrt{2}} \right)^2 \left| V_{u_i b} V_{u_i s}^* \right|^2 
\\ & \times \left( \frac{M_B^2 - M_K^2}{m_b - m_s} \right)^2 \frac{\lambda_{B \to K \phi}^{1/2}}{M_B^3} \left[ f_0^{K} (m_{\phi}^2) \right]^2 \, ,
\label{eq:BRBtoKphi}
\end{align}
where $\lambda_{X \to Y Z} = M_X^4 + M_Y^4 + m_Z^4 - 2 M_X^2 M_Y^2 - 2 M_Y^2 m_Z^2 - 2 m_Z^2 M_X^2$, and $\Gamma_B$ is the total decay width of $B$ meson. 

In this paper, we consider the inclusive bound on the coupling $g_R^B$. 
For $b \to s \phi$ decays, the dominant decay mode is $B^+ \to \bar{c} X$ whose branching ratio is $97 \pm 4 \%$. 
Therefore, we can obtain the inclusive upper bound on any decay mode associated with $b \to s \phi$ to be
\begin{align}
{\rm BR}(b \to s \phi) < 1 - {\rm BR}(B^+ \to \bar{c} X) \lesssim 11 \% \, .
\end{align}

The above bound is independent of the dilaton decay modes. However, considering branching ratios similar to a Higgs-portal scalar, a more stringent bound can be obtained from $\phi$ mediated semileptonic decay channels for the $B$ meson, such as $B \to K + \mu^+ \mu^-$ from LHCb~\cite{LHCb:2016awg}. Following ref.~\cite{Winkler:2018qyg}, this constraint is depicted as the red-shaded region in Fig.~\ref{fig:dilaton_plot}, while the masked regions correspond to charmonium resonances. We reiterate that this bound is dependent on $BR(\phi \to \mu^+ \mu^-)$, and therefore model dependent. For example, if $c_{\phi \mu \mu} \ll 1$, then this branching ratio is suppressed as $c_{\phi \mu \mu}^2$ as the hadronic decay channels do not depend on $c_{\phi \mu \mu}$. Also, if in some specific model, the dilaton primarily decays to a dark sector, then this bound is also relaxed. Note that, even these scenarios can not evade the inclusive $b \to s \phi$ bound we provide in this work. 

Similarly, we also have a constraint from the inclusive $s$ decay. 
The diagrams relevant to $K$ decays are the same as in Fig.~\ref{fig:btosphi} by replacing initial $b \to s$ and final $s \to d$. 
Then, the effective coupling can be estimated from Eq.~\eqref{eq:btosphi} as
\begin{align}
\mathcal{L}_{\rm eff} \supset \frac{4 G_F}{\sqrt{2}} V_{u_i s} V_{u_i d}^* g_R^K \bar{d}_L s_R \phi + {\rm h.c.} \, , \label{eq:stodphi}
\end{align}
where we have omitted the $g_L^K$ term due to the suppression of $m_d / m_s \approx 1/20$, as in the case for $g_{R, L}^B$. 
Using $g_R^K$, the branching ratio for $K \to \pi \phi$ decay can be estimated in a similar manner. For this decay mode, we also have an inclusive bound, and it can be read as
\begin{align}
{\rm BR}(s \to d \phi) < \mathcal{O}(0.1) \% \, .
\end{align}
However, this is a much weaker constraint as compared to the inclusive $b\to s \phi$ bound, and therefore, we do not show it in Fig.~\ref{fig:dilaton_plot}.

Below the threshold mass for the dilaton to decay into two muons, depending on $f_\phi$, the dilaton may decay into two photons, electron-positron pair, or effectively have a decay length long enough such that it appears as an invisible mode in a particular experiment. In this mass range, for the invisible decay signature, the most stringent bound comes from the NA62 experiment which looks for $K^+ \to \pi^+ \nu \bar\nu$. For a vanilla Higgs-portal scalar, this NA62 bound ~\cite{NA62:2021zjw} is the strongest till the threshold $m_K-m_\pi$. However, the dilaton can have an enhanced coupling to photons due to CFT loops, and therefore, can have a decay length much shorter compared to the Higgs-portal scalar. To illustrate, given the length $L_{\rm exp}$, and typical energy $E_{\rm exp}$ of a particular experiment, the fraction of $\phi$ surviving after traversing a length $L_{\rm exp}$ is 
\begin{align}
    f_{\rm sur, exp}&= e^{-\frac{L_{\rm exp}}{c \tau_{\phi} \beta_{\phi} \gamma_{\phi}}} \ ,
\end{align}
where $\tau_\phi$ is the lifetime of $\phi$, and the boost factor $\beta_\phi \gamma_\phi \sim E_{\rm exp}/m_\phi$. Therefore, the NA62 bound on the branching ratio is translated into the parameter space of the dilaton taking into account the exponential tail. Specifically, we demand that
\begin{equation}
    {\rm BR}(K \to \pi \phi) \  f_{\rm sur, NA62} <{\rm BR} (K \to \pi + {\rm inv.})_{\rm NA62} \ ,
\end{equation}
where the left hand side is calculated from the dilaton interaction Lagrangian and the right hand side is the experimental constraint taken from the NA62 experiment~\cite{NA62:2021zjw}. We used the detector length $L_{\rm NA62} = 150$ m, and the energy relevant $E_{\rm NA62}=30$ GeV. In Fig.~\ref{fig:dilaton_plot} this is depicted as a dark yellow region. Note that, for a sufficiently low $f_\phi$, the decay signature of $\phi$ is visible and the NA62 bound is lifted. This is in sharp contrast to the Higgs-portal case. 

On the other hand, if $\phi$ decays visibly, then experimental bounds from $K_L \to \pi^0 \gamma \gamma$ and $K_L \to \pi^0 e^+ e^-$ need to be satisfied. We take the KTeV bounds for these modes~\cite{KTeV:2003sls, KTeV:2008nqz}. As the dilaton is CP even, these bounds are applied without any CP violating suppression factor as opposed to axions. In particular, although $K_L \to \pi^0 \gamma \gamma$ leaves a viable window below the muon production threshold, the stringent $K_L \to \pi^0 e^+ e^-$ constraint nearly closes that window for $c_{\phi e e}=1$. For the KTeV experiment, we used $L_{\rm KTeV} = 1$ m, while the energy relevant was set to $E_{\rm KTeV}=40$ GeV. It is evident that these visible modes are complementary to the invisible modes, and cover the opposite direction in the $f_\phi$ axis as a larger $f_\phi$ would relax the NA62 bound, but the KTeV bounds will become more severe in that case. For mapping the KTeV constraint, we convoluted $K \to \pi \phi$ branching ratio from the theory with BR$(\phi \to \gamma \gamma) (1-f_{\rm sur, KTeV})$, and BR$(\phi \to e^+ e^-) (1-f_{\rm sur, KTeV})$ for the aforementioned modes, respectively.

Note, however, that depending on the composite nature of the electron in the UV theory, this is expected to be different from 1, as alluded to in section~\ref{sec:model}. To be concrete, in the warped extra dimension scenario, this depends on where the electron is localized, and its coupling to the radion/dilaton can be easily suppressed by localizing it towards the UV brane~\cite{Csaki:2020zqz}. Therefore, we also show a contour for $c_{\phi e e}=0.1$ in Fig.~\ref{fig:dilaton_plot}, which revives the low-mass window. We will choose a benchmark point in this region to illustrate a technique of model-independent CP property extraction in the next section.

\subsection{Collider bounds}
\label{sec:LEP}
The primary collider bound on the dilaton comes from LEP searches for light Higgs-like particles in the channel $e^+ e^- \to Z^* \phi$. The bound assumes that the $\phi$ coupling is strictly proportional to the SM Higgs coupling, and hence below the threshold to produce two muons, constraints from invisible mode searches at L3 are applied. Then, if the model dependence of the exactly Higgs-like coupling to the muon is relaxed, still the bound from the invisible mode will be applied~\cite{L3:1996ome}. Similar bounds from CMS and LHCb searches above the $B$-meson mass are weaker than the LEP bound.

A possible search for $e^+ e^- \to \gamma (\phi \to \tau^+ \tau^-)$ can be performed above the $\tau$ production threshold at Belle II. However, the $Z$ boson enjoys a tree-level coupling to the dilaton that is proportional to its mass, and although the increase in luminosity at Belle II somewhat compensates for that, the inferred bounds are comparable to the existing LEP bound, at least in our estimated analysis.

\subsection{Bounds from astrophysics and cosmology}
\label{sec:BBN}

The parameter space of the light dilaton also receives constraints from the early Universe cosmology. In particular, if the lifetime of the dilaton is larger than about $0.1$s, it will spoil the success of the predicted light element abundances from the BBN due to extra energy injection in the thermal bath. We utilize ref.~\cite{Fradette:2017sdd} for the constraint on the lifetime, together with the dilaton couplings \eqref{eq:dilaton_fermion_coupling}, \eqref{eq:dilaton_photon_coupling} to infer bounds on the parameter space of the dilaton as shown in Fig.~\ref{fig:dilaton_plot}. The enhanced coupling to the photon makes the BBN constraint relaxed compared with the Higgs-portal scalar case~\cite{Winkler:2018qyg}. While the assumption of a comparable branching ratio to a Higgs-portal scalar is made, it is expected to give only a marginal effect on the lifetime bound.

If the dilaton is produced on-shell in a supernova explosion, it may carry away significant energy which may shorten the observed duration of the neutrino pulse emission.
Demanding an upper bound on the energy loss rate due to the dilaton-nucleon scattering,
the constraint can be inferred from SN1987a. We follow ref.~\cite{Krnjaic:2015mbs, Ishizuka:1989ts} for the calculation of the energy loss rate per unit volume $Q_\phi$, and demand that 
\begin{equation}
    P_{\rm esc, SN}  \ Q_\phi V_{\rm SN} \lesssim 10^{53}  \ {\rm erg/s} \ ,
    \label{eq:SN1897a}
\end{equation}
where $V_{\rm SN} = (4 \pi/3) R_{\rm SN}^3$ is the supernova volume, $R_{\rm SN} \sim 10$ km being its radius. $P_{\rm esc, SN}$ denotes the escape probability for the dilaton without decaying or getting reabsorbed inside the supernova. The information of the dilaton lifetime enters into the escape probability as
\begin{equation}
     P_{\rm esc, SN} \simeq e^{-R_{\rm SN}/\gamma_\phi \beta_\phi c \tau_\phi} e^{- R_{\rm SN} Q_\phi/\rho_\phi} \ , 
\end{equation}
where $\gamma_\phi \beta_\phi$ is the boost factor as before and $\rho_\phi$ is the equilibrium energy density for the dilaton at the relevant temperature $T_{\rm SN} \sim 30$ MeV. Our inferred constraint is shown as the brown dashed region in Fig.~\ref{fig:dilaton_plot} as this constraint is only estimated at the order of magnitude level. This constraint resembles the Higgs-portal scalar because, for the relevant parameter space, the small coupling makes both decay lengths much bigger than the supernova radius. 

Finally, let us comment on the light radion/dilaton explanation of the muon $g-2$ anomaly. The green band in Fig.~\ref{fig:dilaton_plot} represents the parameter space where the dilaton contribution to the muon $g-2$ makes $\Delta a_\mu$ within $1\sigma$ deviation from the experimentally measured value. However, this region is excluded from the model-independent $b \to s \phi$ constraint shown as gray shaded region in Fig.~\ref{fig:dilaton_plot} together with the LEP bound in the higher mass window. A possible caveat for the LEP bound is that here the dilaton branching ratio is considered exactly similar to the SM Higgs with the same mass, although marginal modification is expected.

\section{Extraction of CP property}
\label{sec:obs}

 \begin{figure*}[ht]
    \subfloat[\label{fig:diagrams}]{
      \includegraphics[width=0.46\textwidth]{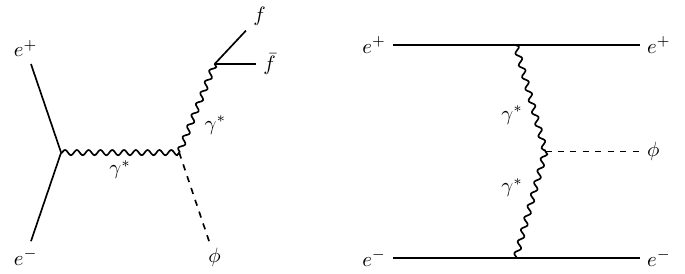}
    }
    \hfill
    \subfloat[\label{fig:pict}]{
      \includegraphics[width=0.46\textwidth]{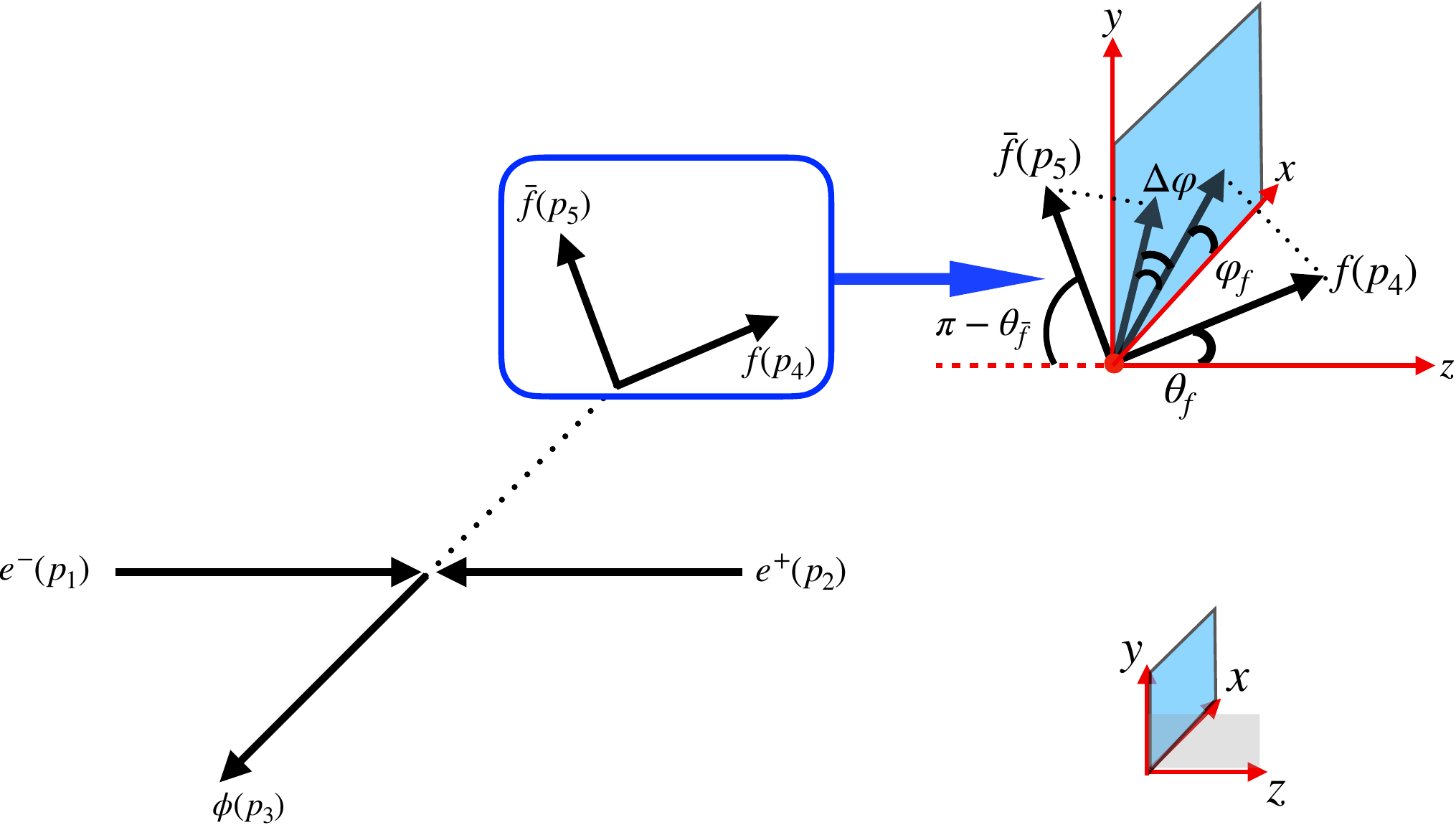}
    }
    \caption{(a) Feynman diagrams for the process $e^+ e^- \to f \bar f \phi$ mediated by a virtual photon. 
The $s$-channel diagram (left) is common for all $f$, while the $t$-channel (right) contributes for $f = e$ only.
(b) a schematic diagram for the momentum of outgoing particles. 
We choose an $e^+ e^-$ beam line to be along with the $z$-axis, while 
the dotted line indicates an off-shell photon $\gamma^*$. The azimuthal angle between the outgoing fermions is denoted as $\Delta \varphi$.}
    \label{fig:collider}
  \end{figure*}

Motivated by the enhanced photon coupling to the dilaton, let us now proceed one step further
in extracting the CP property of a prospective signal due to a light dilaton-mediated process
at a lepton collider like Belle II.
In particular, we will illustrate how to distinguish its signature from a CP-odd axion-like particle (ALP). For a comprehensive exploration of ALPs in the context of rare meson decays see Ref.~\cite{BaBar:2021ich}.
Our method has the merit that it can be applied in a broader scenario even when $\phi$ couples to some dark sector~\cite{Dolan:2017osp}
or decays effectively outside the detector.
Furthermore, the technique can be carried over to any lepton collider machine, as our method is merely kinematical. Following in the footsteps of the earlier works on extracting the CP property of the SM Higgs before its discovery~\cite{Plehn:2001nj}, the strategy relies on the imprints of the CP nature of $\phi$ in the variation of the differential cross-section with the azimuthal angle between the outgoing fermions $\Delta \varphi$. 

To establish our notation for this part, let us denote the effective interaction of the dilaton to the photon as 
\begin{align}
\mathcal{L_{{\rm eff}, \phi \gamma}} = - \frac{g_{\phi \gamma \gamma}}{4} \phi F^{\mu \nu} F_{\mu \nu} \, ,
\label{eq:gphigammagamma}
\end{align}
where $F$ is the electromagnetic field strength tensor.
On the other hand, the coupling of an ALP, or merely an axion, ($a$) to photons can be parametrized as
\begin{align}
\mathcal{L}_{{\rm eff}, a \gamma} = - \frac{g_{a \gamma \gamma}}{4} a F^{\mu \nu} \widetilde{F}_{\mu \nu} \, ,
\end{align}
with the dual field strength $\widetilde{F}_{\mu \nu} \equiv \frac{1}{2} \epsilon_{\mu \nu \rho \sigma} F^{\rho \sigma}$. 
For the following discussion, $\phi$ will generically denote both the CP-odd and CP-even (pseudo)scalars.

\subsection{Imprints of CP in $e^+ e^- \to f \bar{f} \phi$ }
\label{sec:CP}

We consider the dilaton production $e^+ (p_1) e^- (p_2) \to \phi (p_3) f(p_4) \bar f(p_5) $ where $f$ denotes an SM charged lepton and the four-momentum assignments are shown in appendix~\ref{sec:calc}. 
Our focus is on the outgoing lepton case as it is easier to measure their momenta precisely as compared with jets. The Feynman diagram for this process is depicted in Fig.~\ref{fig:diagrams}, where we use the coupling with the photons, namely $g_{\phi \gamma \gamma}$ in Eq.~\eqref{eq:gphigammagamma}, to produce $\phi$ and then the off-shell photon can generate the outgoing lepton pair. The azimuthal angle between the outgoing leptons is the angle of interest $\Delta \varphi$, and is elucidated in the schematic Fig.~\ref{fig:pict}. Concretely,
\begin{equation}
    {\widehat{{p_4}}_{T} \cdot \widehat{{p_5}}_{T}} = \cos (\Delta \varphi) \ ,
    \label{eq:DeltaPhi}
\end{equation}
where $\vec{p_i}_{T}$ is the transverse component of the momentum vector with respect to the beam axis, and the overhat denotes a unit vector. Manifestly, $\Delta \varphi$ is boost invariant along the beam axis.
We note that at a center of mass (COM) energy lower than the electroweak scale, as applicable to Belle II,
the photon-mediated process dominates over the one mediated by the $Z$ boson. We will look at the variation of the normalized differential cross section with $\Delta \varphi$. 

For $f \neq e$, only the $s$-channel diagram contributes (left subfigure in Fig.~\ref{fig:diagrams}),
while for $f = e$ one also has to include the $t$-channel diagram (right subfigure in Fig.~\ref{fig:diagrams}). The total amplitude for $f = e$ is therefore
\begin{align}
\overline{\left| \cal{M} \right|^2} = \overline{\left| \mathcal{M}_s + \mathcal{M}_t \right|^2} = \overline{\left| \mathcal{M}_s \right|^2} + 2 {\rm Re}\overline{\mathcal{M}_s^* \mathcal{M}_t} + \overline{\left| \mathcal{M}_t \right|^2} \, ,
\label{eq:AmpSq} 
\end{align}
while for $f \neq e$,
\begin{align}
\overline{\left| \cal{M} \right|^2} = \overline{\left| \mathcal{M}_s \right|^2} \, .
\label{eq:AmpSqfneqe} 
\end{align}
Here overline indicates the average and sum over the initial and final fermion spins, respectively. 
We summarize the results for each channel and interference term, while the details of the calculation including the phase space integration are presented in appendix~\ref{sec:calc}. 
 \begin{figure*}[t]
   \subfloat[\label{1GeV:fig}]{
      \includegraphics[width=0.46\textwidth]{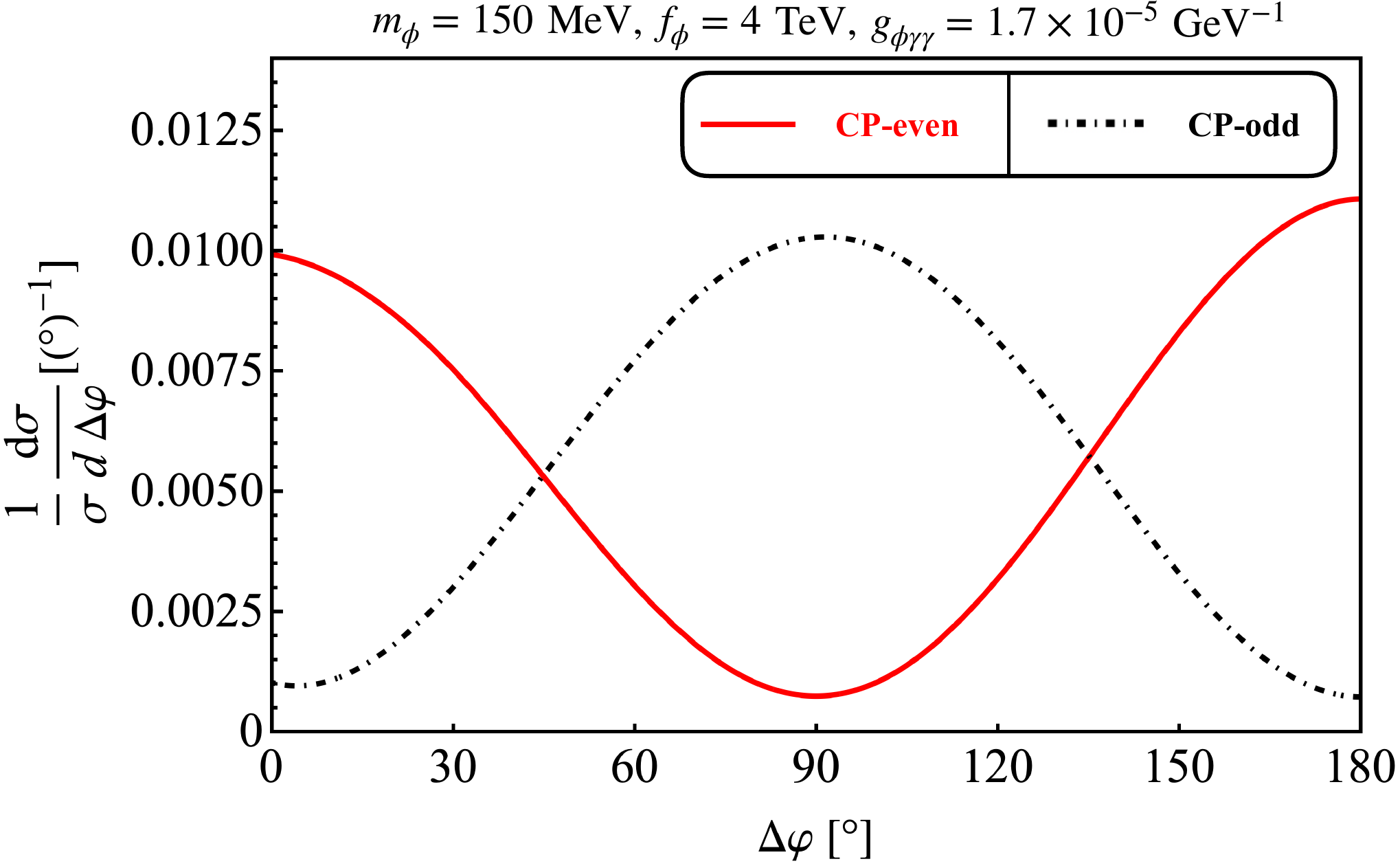}
    }
    \hfill
    \subfloat[\label{secluded:fig}]{
      \includegraphics[width=0.46\textwidth]{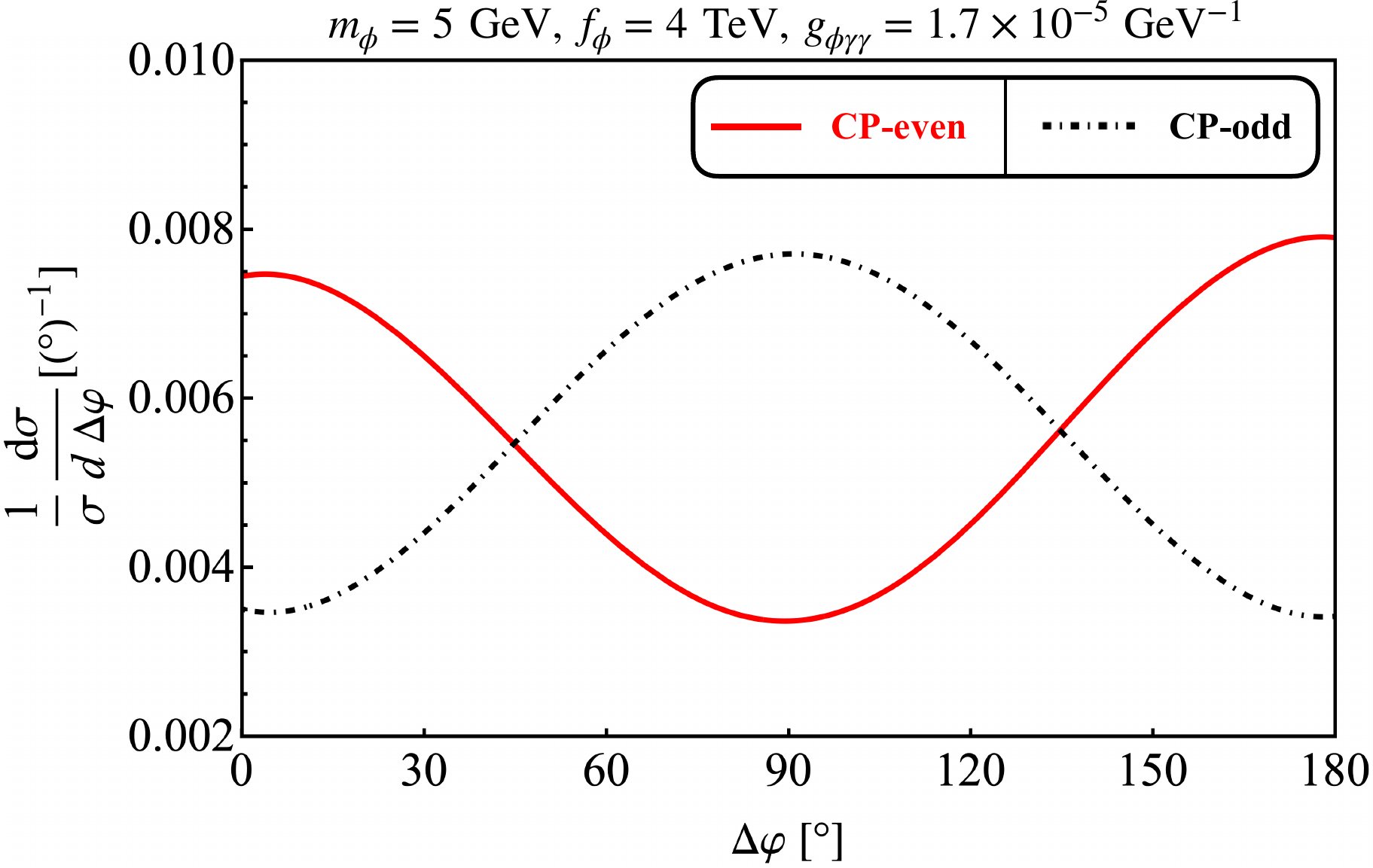}
    }
    \hfill
    \subfloat[\label{decaying:fig}]{
      \includegraphics[width=0.46\textwidth]{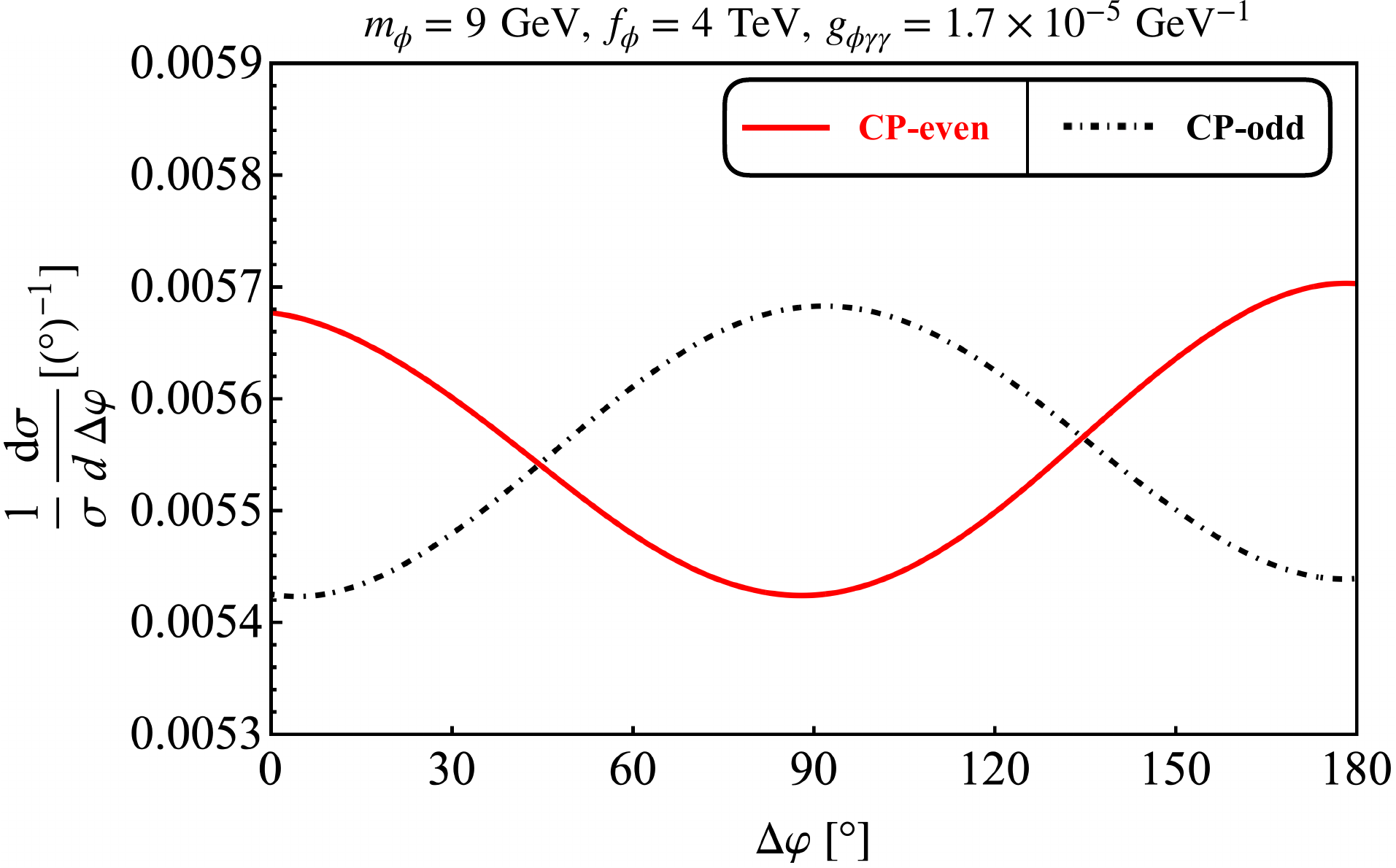}
    }
    \hfill
    \subfloat[\label{muon:fig}]{
      \includegraphics[width=0.46\textwidth]{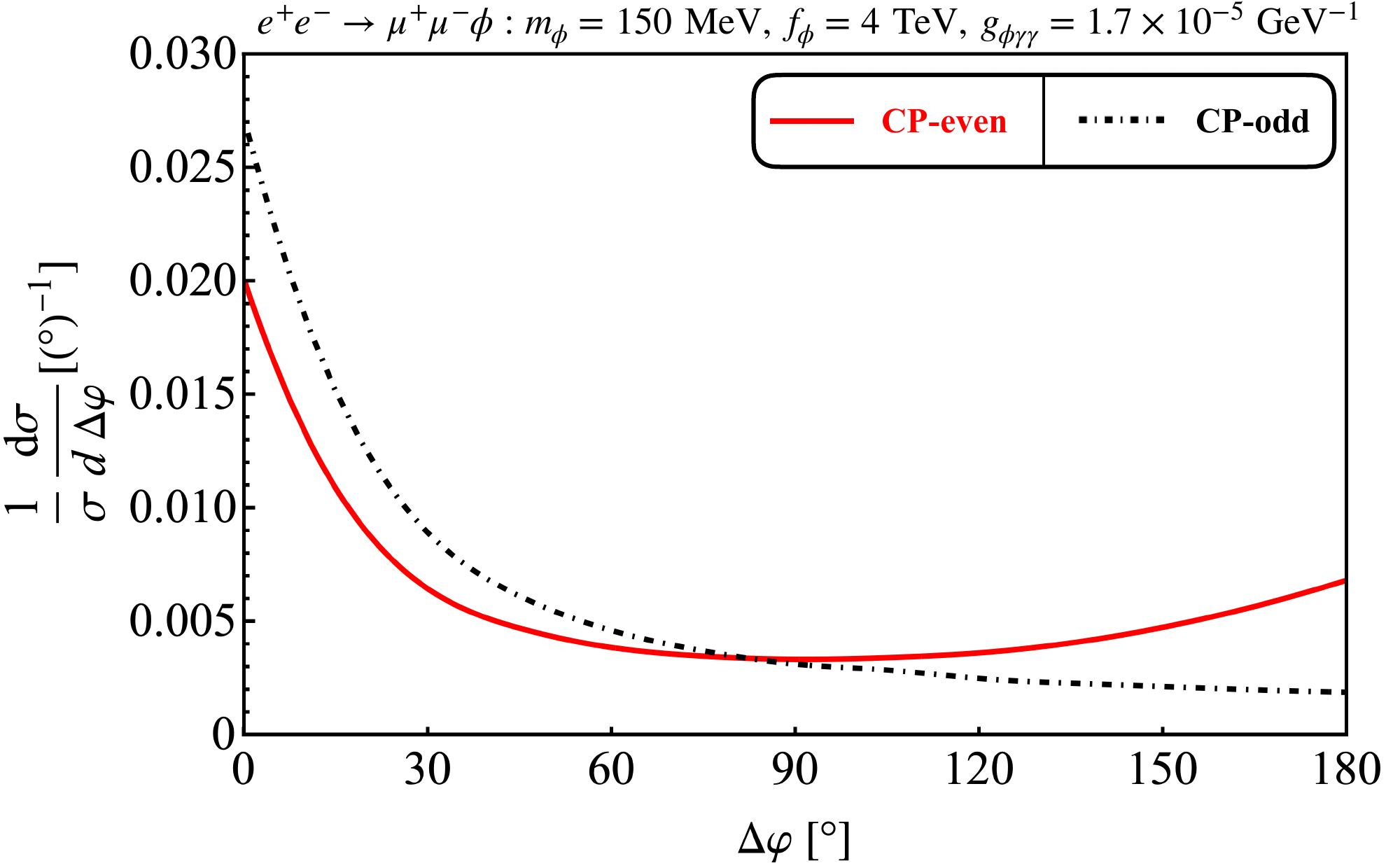}
    }
    \caption{The variation of the normalized differential cross section with respect to azimuthal angle between the outgoing leptons $\vpfb$ for the electron in the final state with $m_{\phi} = 150$ MeV (top left panel), $5$ GeV (top right panel) and $9$ GeV (bottom left panel) and the muon in the final state with $m_{\phi} = 150$ MeV (bottom right panel). The red solid (black dot-dashed) curve represents the CP-even (CP-odd) case.
    }
    \label{fig:plot-mphidif}
  \end{figure*}
The contribution from the $s$-channel in Eq.~\eqref{eq:AmpSq} is 
\begin{align}
\nonumber
\overline{\left| \mathcal{M}_s \right|^2} &\simeq \frac{32 \pi^3 \alpha_{\rm em}^2 \alpha_\phi}{p_{12}^2 p_{45}^2} \Bigl[ 2 \left( p_{14} p_{25} - p_{24} p_{15} \right)^2  
\\ & + p_{12} p_{45} \left\{ \left( p_{14} - p_{25} \right)^2 + \left( p_{24} - p_{15} \right)^2 + 2 p_{12} p_{45}  \right\} \Bigr] \ , \label{eq:Msqs+}
\end{align}
for CP-even $\phi$, and
\begin{align}
\nonumber
\overline{\left| \mathcal{M}_s \right|^2} &\simeq \frac{32 \pi^3 \alpha_{\rm em}^2 \alpha_\phi}{p_{12}^2 p_{45}^2} \Bigl[ - 2 \left( p_{14} p_{25} - p_{24} p_{15} \right)^2 
\\ & + p_{12} p_{45} \left\{ \left( p_{14} + p_{25} \right)^2 + \left( p_{24} + p_{15} \right)^2 - 2 p_{12} p_{45}  \right\} \Bigr] \ , \label{eq:Msqs-}
\end{align}
for CP-odd $\phi$. 
Here, $\alpha_{\rm em}$ denotes the fine-structure constant, and we have defined $p_{ij} \equiv p_i \cdot p_j$ and $\alpha_{\phi} \equiv g_{\phi \gamma \gamma}^2 / (4 \pi)$. 
Note that these expressions are given in the limit of $m_f/{\sqrt{s}}, \, m_e/{\sqrt{s}} \to 0$ for conciseness,
while we take the full expressions without taking the limit into account for our numerical calculations.
$\sqrt{s}$ denotes the COM energy of the collider concerned. The inner products in the denominator are due to the off-shell photon propagators as usual, $q_1^2 = (p_1 + p_2)^2 \simeq 2 p_1 \cdot p_2$ and $q_2^2 = (p_4 + p_5)^2 \simeq 2 p_4 \cdot p_5$. Notice the differences in signs for some of the terms in the above equations are a consequence of the different Lorentz structure of the photon-(pseudo)scalar Lagrangian, which in turn is connected to their CP property.
The $t$-channel contribution in Eq.~\eqref{eq:AmpSq} is
\begin{align}
\nonumber
\overline{\left| \mathcal{M}_t \right|^2} &\simeq \frac{32 \pi^3 \alpha_{\rm em}^2 \alpha_\phi}{p_{14}^2 p_{25}^2} \Bigl[ 2 \left( p_{12} p_{45} - p_{24} p_{15} \right)^2 \\
& + p_{14} p_{25} \left\{ \left( p_{12} - p_{45} \right)^2 + \left( p_{24} - p_{15} \right)^2 + 2 p_{14} p_{25}  \right\} \Bigr] \ , \label{eq:Msqt+}
\end{align}
for CP-even $\phi$, and
\begin{align}
\nonumber
\overline{\left| \mathcal{M}_t \right|^2} &\simeq \frac{32 \pi^3 \alpha_{\rm em}^2 \alpha_\phi}{p_{14}^2 p_{25}^2} \Bigl[ - 2 \left( p_{12} p_{45} - p_{15} p_{24} \right)^2 
\\ & + p_{14} p_{25} \left\{ \left( p_{12} + p_{45} \right)^2 + \left( p_{24} + p_{15} \right)^2 - 2 p_{14} p_{25}  \right\} \Bigr] \ , \label{eq:Msqt-}
\end{align}
for CP-odd $\phi$. 
For this channel, the (inverse of) off-shell photon propagators are $(q'_1)^2 = (p_1 - p_4)^2 \simeq - 2 p_1 \cdot p_4$ and $(q'_2)^2 = (p_2 - p_5)^2 \simeq - 2 p_2 \cdot p_5$. 
Note that these expressions can be obtained from the $s$-channel results by replacing $p_2 \leftrightarrow p_4$, namely, $p_{12} \leftrightarrow p_{14}$, $p_{25} \leftrightarrow p_{45}$. 
The interference term is
\begin{align}
\nonumber
\overline{\mathcal{M}_s^* \mathcal{M}_t} &\simeq \frac{16 \pi^3 \alpha_{\rm em}^2 \alpha_\phi}{p_{12} p_{14} p_{25} p_{45}} \Bigl[ 2 \left( p_{12} p_{45} - p_{14} p_{25} \right)^2 
 + \left( p_{24} - p_{15} \right)^2 \\ & \times \left( p_{12} p_{45} + p_{14} p_{25} \right) - p_{15} p_{24} \left( p_{15}^2 + p_{24}^2 \right) \Bigr] \ , \label{eq:Msqts+}
\end{align}
for CP-even $\phi$, and
\begin{align}
\nonumber
\overline{\mathcal{M}_s^* \mathcal{M}_t} &\simeq \frac{16 \pi^3 \alpha_{\rm em}^2 \alpha_\phi}{p_{12} p_{14} p_{25} p_{45}} \Bigl[ - 2 \left( p_{12} p_{45} - p_{14} p_{25} \right)^2 
\\ & \nonumber  + \left( p_{24} + p_{15} \right)^2 \left( p_{12} p_{45} + p_{14} p_{25} \right) \\ & - p_{15} p_{24} \left( p_{15}^2 + p_{24}^2 \right) \Bigr] \ , \label{eq:Msqts-}
\end{align}
for CP-odd $\phi$. We note that the $t$-channel cross section can be larger than the $s$-channel one because we can take small $(q'_{1, 2})^2$ for two off-shell photons. 
On the other hand, for the $s$-channel process, we cannot take small $q_1^2$ as it is determined by the COM energy. 

For the purpose of illustrating the distinct CP property, let us choose a few benchmark points
from the allowed parameter space in Fig.~\ref{fig:dilaton_plot}.
We take $m_\phi = 0.15, 5, 9$ GeV with $f_\phi = 4$ TeV.
For $\Lambda_{\rm UV}=10^{10}$ GeV, using Eq.~\eqref{eq:dilaton_photon_coupling},
the inferred value of $g_{\phi \gamma \gamma} \sim 1.7 \times 10^{-5}$ GeV$^{-1}$,
which also satisfies constraints on the (pseudo)scalar-photon coupling~\cite{Bauer:2021mvw}.
This choice allows us to portray the behavior with the dilaton mass for three distinct regions
for a collider like Belle II with $\sqrt{s} \sim 10$ GeV.

\subsection{Angular distribution of differential cross section}
\label{sec:efinal}

In Fig.~\ref{fig:plot-mphidif}, we show the result of the variation of the normalized differential cross section with respect to $\vpfb$. 
 \begin{figure*}[t]
   \subfloat[\label{5GeVwBG:fig}]{
      \includegraphics[width=0.46\textwidth]{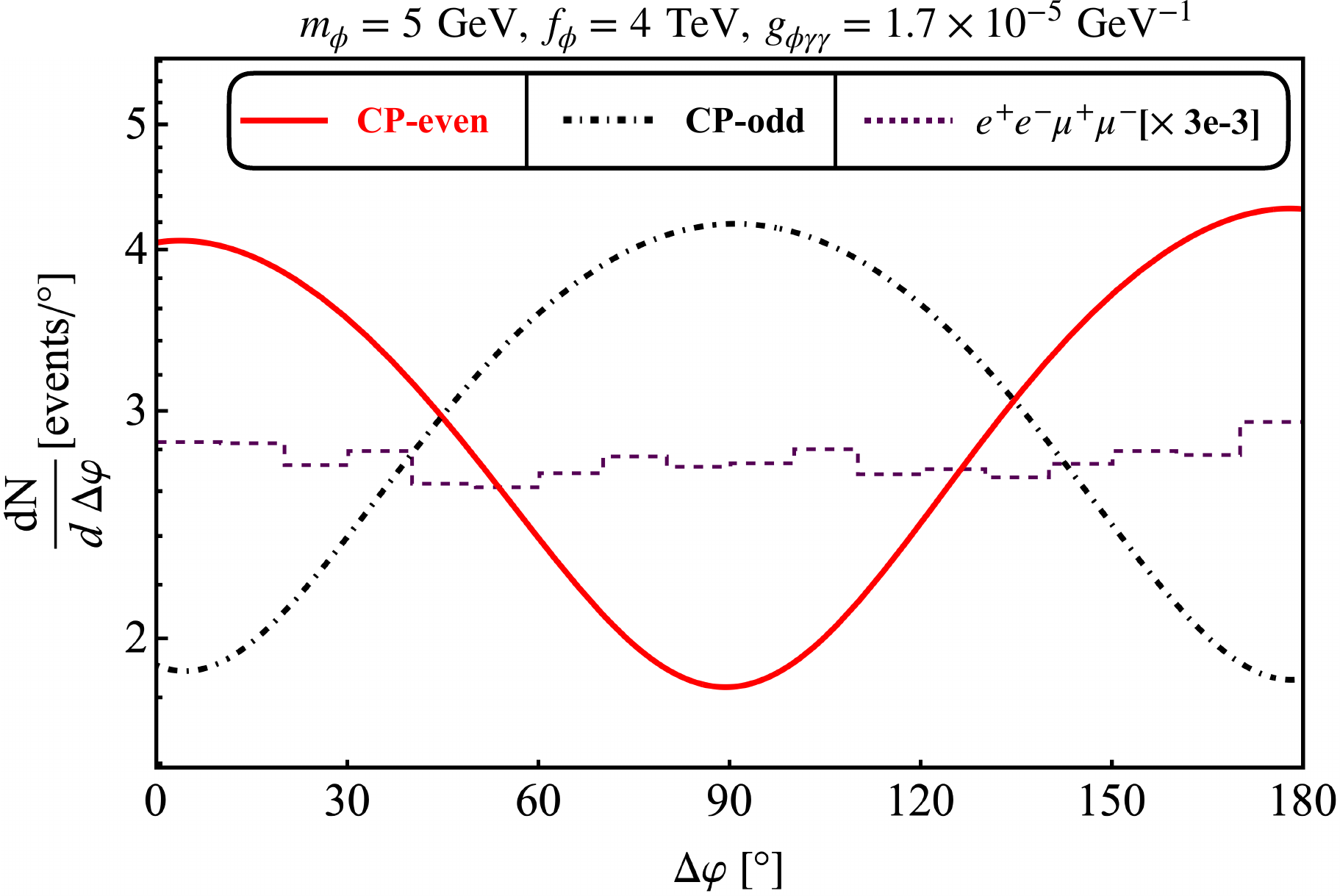}
    }
    \hfill
    \subfloat[\label{5GeVwBGTa:fig}]{
      \includegraphics[width=0.46\textwidth]{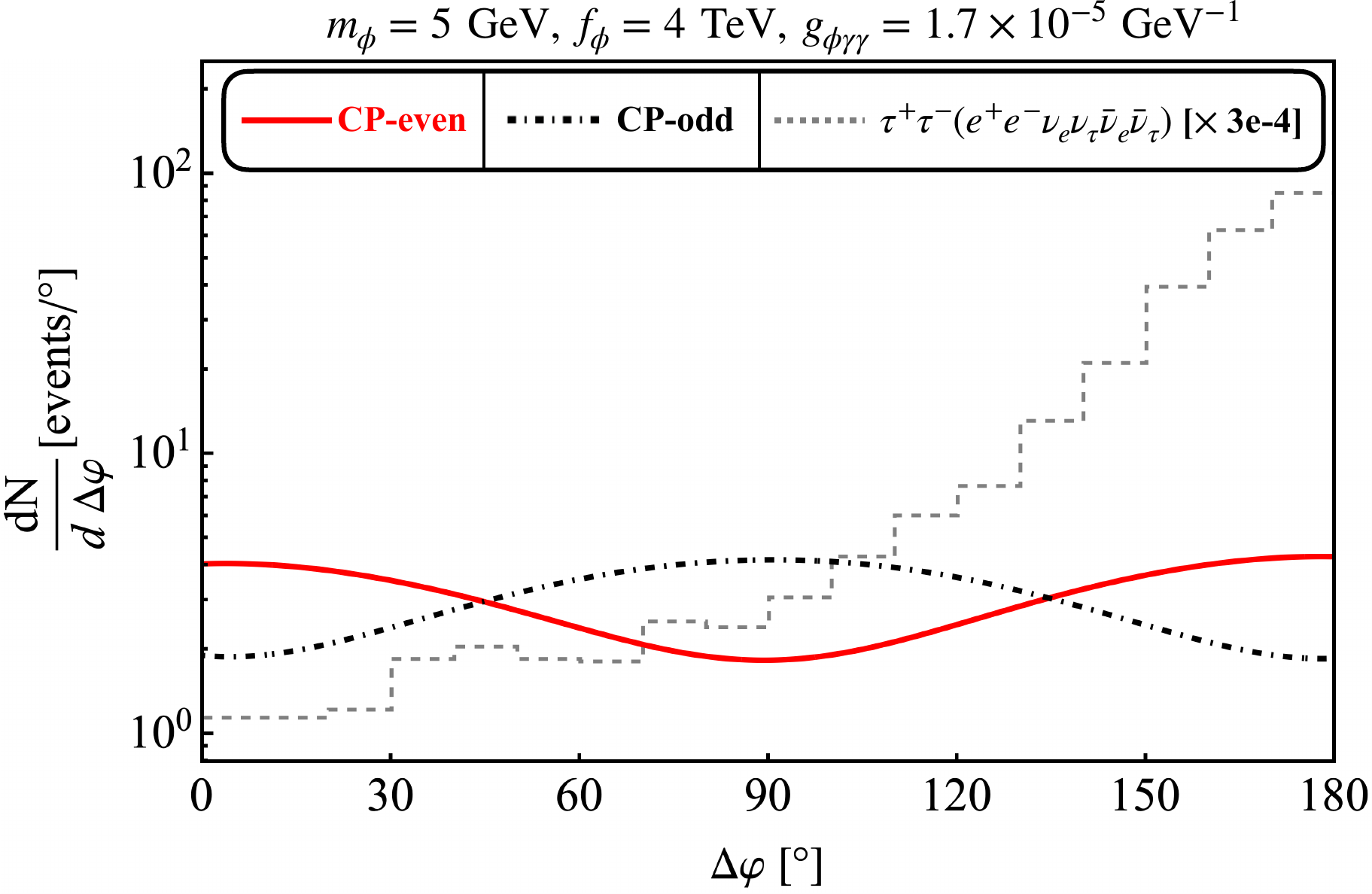}
    }
    \caption{Variation of the expected number of signal and principal background events with respect to the azimuthal angle between the outgoing $e^+ e^-$ pair for a representative benchmark of $m_\phi=5$ GeV, where we have assumed the Belle II integrated luminosity as $50$ab$^{-1}$. 
    (a) We consider visible decay mode $\phi \to \mu^+ \mu^-$ and hence show the SM background $e^+ e^- \to e^+ e^- \mu^+ \mu^-$ (purple dashed), scaled down by $3 \times 10^{-3}$. A lower bound $m_{ee} > 0.1$ GeV was employed in order to trim down the collinear peak towards low $\Delta \varphi$. This background profile is flat with $\Delta \varphi$, aiding with the discrimination of the CP-even (red solid) from the CP-odd (black dot-dashed) cases. 
    (b) We consider invisible decay mode for $\phi$ with the same mass and show the SM background (gray dashed) from $e^+ e^- \to \tau^+ \tau^-$ where the taus decay to electrons. This is peaked towards large $\Delta \varphi$ and we have scaled the background by $3 \times 10^{-4}$.
    }
    \label{fig:5GeVwBG}
  \end{figure*}
For the case with the electron in the final state, the distinct signatures of CP-odd and CP-even cases are evident. 
Apart from this angular difference, the total cross sections are similar for both cases.
Numerically, for $m_\phi \ll \sqrt{s}$, we find
\begin{align}
\sigma_{e^+ e^- \to e^+ e^- \phi} \simeq 50 \times \left( \frac{g_{\phi \gamma \gamma}}{10^{-5}} \right)^2 ~ {\rm ab} \, .
\end{align}
Let us comment on the specific case of Belle II as an application. Their final integrated luminosity goal is $50 {\rm ab}^{-1}$, and hence, we can expect $\sim 2500$ events in total.
It should be emphasized that for the Higgs portal case,
with the same given $f_\phi$, we get $g_{\phi \gamma \gamma} \sim 10^{-6}$ GeV$^{-1}$~\cite{Marciano:2011gm},
and expect $\sim 25$ events, which is two orders of magnitude below the number of events for the dilaton case, and not enough for extracting the CP property. 

The qualitative behavior of Fig.~\ref{fig:plot-mphidif} can be understood as follows. 
For the CP-odd pseudoscalar, the amplitude is proportional to the Levi-Civita tensor, and therefore,
four distinct momenta are required to contract all Lorentz indices.\footnote{
Although the squared amplitude has $g^{\mu \nu}$ term as well as $p_i^{\mu} p_j^{\nu}$,
its contribution is smaller than the $p_i^{\mu} p_j^{\nu}$ terms with $\vpfb \neq 0, \pi$. This is evident in Fig.~\ref{fig:plot-mphidif}.}
Therefore, the amplitude will diminish if the transverse components of $p_4$ and $p_5$ are parallel, which is realized with $\vpfb = 0$ or $\pi$. 
For the CP-even scalar, the inner product of $p_4 \cdot p_5$ has an important role in the behavior. 
If we focus on the inner product of transverse components, we have
\begin{align}
\vec{p_4}_T \cdot \vec{p_5}_T = \sqrt{E_4^2 - m_f^2} \sqrt{E_5^2 - m_f^2} s_{\thf} s_{\thfb} c_{\vpfb} \, ,
\label{eq:p4Tp5T}
\end{align}
and hence the squared amplitude will decrease around $\vpfb = \pi / 2$. Here $E_4$, $E_5$, $\theta_f$, $\theta_{\bar f}$ are the energies and polar angles of the outgoing fermions, and are illustrated in Fig.~\ref{fig:pict}. See appendix~\ref{sec:kine} for further details.
Our results agree with ref.~\cite{Plehn:2001nj} qualitatively. 

From Fig.~\ref{fig:plot-mphidif} it is clear that the relative difference of the CP-odd vs.~CP-even cases decreases with an increase in $m_\phi$ due to the reduction of the available phase space.
Besides, the total cross section also decreases
\begin{align}
\sigma_{e^+ e^- \to e^+ e^- \phi} \simeq
\begin{cases}
{\displaystyle 6 \times \left( \frac{g_{\phi \gamma \gamma}}{10^{-5}} \right)^2 ~ {\rm ab}} &  \text{($m_{\phi} = 5$ GeV)} \, , \\[2.0ex]
{\displaystyle 0.4 \times \left( \frac{g_{\phi \gamma \gamma}}{10^{-5}} \right)^2 ~ {\rm ab}} &  \text{($m_{\phi} = 9$ GeV)} \, .
\end{cases}
\end{align}
Therefore, it is easier to distinguish the CP property for the lighter mass.

For nearly all the range of $\vpfb$, the differential cross section is dominated by the $t$-channel contribution,
due to the fact that both (the inverse of) propagators of the virtual photon can be small, $(q'_{1, 2})^2 \sim 0$. 
For $s$-channel, on the other hand, only $q_2^2$ can be small for $\vpfb \sim 0$. 
We can check this directly as
\begin{align}
q_2^2 &= p_4^2 + 2 p_4 \cdot p_5 + p_5^2 \nonumber \\[0.5ex]
&= 2 \left[ m_e^2 + E_4 E_5 - |\vec{p_4}| |\vec{p_5}| \left( c_{\thf}^2 + s_{\thf}^2 c_{\vpfb} \right) \right] \nonumber \\[0.5ex]
&\to 2 \left( m_e^2 + E_4 E_5 - |\vec{p_4}| |\vec{p_5}| \right) \quad (\vpfb \to 0) \, .
\end{align}
If $|\vec{p_4}| |\vec{p_5}| \approx E_4 E_5$, we obtain $q_2^2 \approx 2 m_e^2 \ll 2 E^2$, and hence, the $s$-channel contribution is enhanced. 

Next, we discuss about the process $e^+ e^- \to \phi \mu^+ \mu^-$. 
The result is shown in the bottom right panel of Fig.~\ref{fig:plot-mphidif}. 
Here, we set { $m_{\phi} = 150$ MeV} with a fixed $f_\phi$. 
It is found that for the range of $\pi/2 \leq \vpfb < \pi$, the distributions for the CP-even and CP-odd cases
are slightly different: increasing for the CP-even case, while decreasing for the CP-odd case. 
However, this might not be enough to distinguish the CP property of $\phi$, compared with the electron case. 
This is because, for the process with the muon in the final state, we have $s$-channel only, and the total cross section is much smaller than the electron case, due to no enhancement from $t$-channel. 
We obtain
\begin{align}
\sigma_{e^+ e^- \to \mu^+ \mu^- \phi} \simeq 0.02 \times \left( \frac{g_{\phi \gamma \gamma}}{10^{-5}} \right)^2 ~ {\rm ab} \,  \ \  (m_\phi = 0.15 \  {\rm GeV}) \, ,
\end{align}
both for the CP-even and CP-odd cases.
Therefore, we conclude that $e^+ e^- \to \phi e^+ e^-$ will be the ideal process for our kinematical method
to distinguish the CP property. 

A detailed analysis with the Belle II detector setup is left for a future study
while we comment on the possible background for the $e^+e^-\phi$ mode briefly. 
The dilaton heavier than the dimuon threshold can decay to dimuon, and then we consider the $e^+e^- \mu^+\mu^-$ background for this signal with MadGraph 5~\cite{Alwall:2014hca}. After removing the collinear peak by placing a lower bound on the $e^+e^-$ invariant mass, $m_{ee}>0.1$\,GeV, we consider the bin size of 80\,MeV for the dimuon invariant mass inferred by the detector resolution of the signal peak. The expected yield of the background is 2$\times 10^4$, which is not significant even before the detailed selections. Furthermore, the trimmed resulting background distribution is flat with respect to $\Delta \varphi$, which is distinct from the signal distribution. This background is shown as the purple dashed line in Fig.~\ref{5GeVwBG:fig}, where we have scaled it by $3 \times 10^{-3}$. Therefore, the difference between CP-odd and CP-even particles is expected to be seen in this channel. 

When the dilaton is effectively invisible, the potential background arises either from undetected photons or from the production of a tau lepton pair that decays to $e^+e^-$, associated with neutrinos. In a benchmark with the lowest mass $m_\phi=150$\, MeV, the background estimate studied for the muonic force mediator can be applied \cite{Jho:2019cxq}.  Assuming the background structure of $\mu^+\mu^-+$inv is similar for  $e^+e^-+$inv,  almost all the background can be removed. For the higher mass $m_\phi=5,9$ GeV, the dilaton signature becomes invisible if it dominantly decays into some dark sector particles. We find that the undetected photon background that peaks at the zero missing mass is negligible, but the tau lepton background is sizable, particularly in large $\Delta\varphi$, based on MadGraph 5. The background is selected to have missing mass around the dilaton mass up to the detector resolution of 560(110)\,MeV for $m_\phi=5(9)$\,GeV, and the background shape is included in Fig.~\ref{5GeVwBGTa:fig} as the gray dashed line, where we have scaled it by $3 \times 10^{-4}$.  Dedicated analysis, like reconstructing the tau lepton pair using the collinear approximation, could reduce this type of background.

\section{Conclusions}
\label{sec:sum}

The existence of approximate scale invariance with dynamical spontaneous breaking is a promising candidate of
physics beyond the SM to address the electroweak naturalness problem,
and the dilaton is the key probe into such scenario for a low-energy observer.
In the present paper, we have analyzed the constraints on a light dilaton with a mass spanning the MeV-GeV range,
imposed by experimental data primarily from rare meson decays.
We provided a new inclusive bound from the $b \to s \phi$ transition.
This bound, together with the collider bounds in the higher mass window, elucidates
the status of the light dilaton explanation of the muon $g-2$ anomaly.
It has been shown that despite the parallels between the dilaton and a Higgs-portal scalar, the dilaton coupling to the photon is substantially enhanced due to the involvement of loops from the conformal sector. Consequently, the dilaton's reduced lifetime partly mitigates constraints from searches for $K \to \pi$ + invisible processes at the NA62 experiment and from considerations related to BBN. By capitalizing on this distinctive aspect, we have developed a strategy for extracting the CP property of the dilaton at a lepton collider, such as the ongoing Belle II experiment,
using the variation of the differential cross-section of $e^+ e^- \to e^+ e^- \phi$ with the azimuthal angle
between the outgoing leptons.

\section*{Acknowledgements}

YN is supported by the Natural Science Foundation of China under grant No. 12150610465.
KT is supported by in part the US Department of Energy grant DE-SC0010102 and JSPS Grant-in-Aid for Scientific Research (Grant No. 21H01086).

\appendix

\section{Details of the calculations}
\label{sec:calc}

\subsection{Kinematics}
\label{sec:kine}

In the discussion of section~\ref{sec:obs}, we assign the four momenta as follows:
\begin{align}
e^- &: p_1 = (E, 0, 0, p_z) \, , \label{eq:p1CoM} \\[0.5ex]
e^+ &: p_2 = (E, 0, 0, - p_z) \, , \label{eq:p2CoM} \\[0.5ex]
f &: p_4 = (E_4, |\vec{p_4}| s_{\thf} c_{\vpf}, |\vec{p_4}| s_{\thf} s_{\vpf}, |\vec{p_4}| c_{\thf}) \, , \label{eq:p4CoM} \\[0.5ex]
\bar{f} &: p_5 = (E_5, |\vec{p_5}| s_{\thfb} c_{\vpf + \vpfb}, |\vec{p_5}| s_{\thfb} s_{\vpf + \vpfb}, |\vec{p_5}| c_{\thfb}) \, , \label{eq:p5CoM}
\end{align}
where we consider the center of mass (COM) frame of $e^+ e^-$. 
Here, we use $s_{\alpha} \equiv \sin \alpha$ and $c_{\alpha} \equiv \cos \alpha$.
For the azimuth angle of $p_5$, we define $\vpfb$ as a deviation from $\vpf$. 
Note that we set the $z$-axis along with the $e^+ e^-$ beam line, and
use the four-momentum conservation to parameterize the momentum of $\phi$,
which is denoted as $p_3$, by the other parameters:
\begin{align}
p_3^{\mu} = p_1^{\mu} + p_2^{\mu} - p_4^{\mu} - p_5^{\mu} \, .
\label{eq:4momconserve}
\end{align}
The parameterizations of angles in $p_4$ and $p_5$ are defined for later convenience. 
The other constraints come from the $p^2 = m^2$ relations, which are
\begin{align}
(p_1)^2 = m_e^2 ~~ &\Rightarrow ~~ p_z = \sqrt{E^2 - m_e^2} \approx E \, , \label{eq:p1me} \\[0.5ex]
(p_4)^2 = m_f^2 ~~ &\Rightarrow ~~ |\vec{p}_4| = \sqrt{E_4^2 - m_f^2} \, , \label{eq:p4mf} \\[0.5ex]
(p_5)^2 = m_f^2 ~~ &\Rightarrow ~~ |\vec{p}_5| = \sqrt{E_5^2 - m_f^2} \, , \label{eq:p5mf}
\end{align}
where we take a limit of $m_e^2 / E^2 \to 0$. 
Note that for $m_f$, it is not trivial to take a limit of $m_f^2 / E_{4,5}^2 \to 0$,
because $E_{4,5}$ can be as small as the $m_f$ scale. 
Still, the limit $m_f^2 / E^2 \to 0$ can be justified for $f = e, \mu$ and other light fermions. 

Now, we can calculate all the inner products straightforwardly, and the results can be read as
\begin{align}
\indot{p_1}{p_2} &= 2 E^2 - m_e^2 \approx 2 E^2 \, , \label{eq:p1p2com} \\[0.5ex]
\indot{p_1}{p_3} &\approx E \Big( 2 E - E_4 - E_5  + \sqrt{E_4^2 - m_f^2} c_{\thf} \nonumber \\ & {\hspace{4cm}}  + \sqrt{E_5^2 - m_f^2} c_{\thfb} \Big) \, , \label{eq:p1p3com} \\[0.5ex]
\indot{p_1}{p_4} &\approx E \left( E_4 - \sqrt{E_4^2 - m_f^2} c_{\thf} \right) \, , \label{eq:p1p4com} \\[0.5ex]
\indot{p_1}{p_5} &\approx E \left( E_5 - \sqrt{E_5^2 - m_f^2} c_{\thfb} \right) \, , \label{eq:p1p5com} \\[1.0ex]
\indot{p_2}{p_3} &\approx E \Big( 2 E - E_4 - E_5  - \sqrt{E_4^2 - m_f^2} c_{\thf} \nonumber \\ & {\hspace{4cm}} -\sqrt{E_5^2 - m_f^2} c_{\thfb} \Big) \, , \label{eq:p2p3com} \\[0.5ex]
\indot{p_2}{p_4} &\approx E \left( E_4 + \sqrt{E_4^2 - m_f^2} c_{\thf} \right) \, , \label{eq:p2p4com} \\[0.5ex]
\indot{p_2}{p_5} &\approx E \left( E_5 + \sqrt{E_5^2 - m_f^2} c_{\thfb} \right) \, , \label{eq:p2p5com} \\[1.0ex]
\indot{p_3}{p_4} &= (2 E - E_5) E_4 - m_f^2 + \sqrt{E_4^2 - m_f^2} \sqrt{E_5^2 - m_f^2} f_{\theta} \, , \label{eq:p3p4com} \\[0.5ex]
\indot{p_3}{p_5} &= (2 E - E_4) E_5 - m_f^2 + \sqrt{E_4^2 - m_f^2} \sqrt{E_5^2 - m_f^2} f_{\theta} \, , \label{eq:p3p5com} \\[1.0ex]
\indot{p_4}{p_5} &= E_4 E_5 - \sqrt{E_4^2 - m_f^2} \sqrt{E_5^2 - m_f^2} f_{\theta} \, , \label{eq:p4p5comexplicit}
\end{align}
where ``$\approx$" corresponds to the limit $m_e^2 / E^2 \to 0$ as in Eq.~\eqref{eq:p1me}, and we define
\begin{align}
f_{\theta} \equiv s_{\thf} s_{\thfb} c_{\vpfb} + c_{\thf} c_{\thfb} \, .
\label{eq:fthetadef}
\end{align}
Note that Eq.~\eqref{eq:p4p5comexplicit} can be rewritten by using the four-momentum conservation of Eq.~\eqref{eq:4momconserve} and Eqs.~\eqref{eq:p4mf}, \eqref{eq:p5mf}:
\begin{align}
&(p_4 + p_5)^2 = (p_1 + p_2 - p_3)^2  = 4 E^2 - 4 E E_3 + m_{\phi}^2 \, , \nonumber \\[0.7ex]
\text{and} ~~ &(p_4 + p_5)^2 = p_4^2 + 2 \indot{p_4}{p_5} + p_5^2 = 2 m_f^2 + 2 \indot{p_4}{p_5} \, , \nonumber
\end{align}
where we use $(p_3)^2 = m_{\phi}^2$ and $E_3 = \sqrt{|\vec{p_3}|^2 + m_{\phi}^2}$. 
Comparing these two relations, we obtain
\begin{align}
\indot{p_4}{p_5} &= \frac{1}{2} \left( - 4 E^2 + 4 E E_4 + 4 E E_5 + m_{\phi}^2 \right) - m_f^2 \, . \label{eq:p4p5com}
\end{align}
Eq.~\eqref{eq:p4p5comexplicit} and Eq.~\eqref{eq:p4p5com} should be the same, and hence we can obtain the solution for one of $(E_4, E_5)$ (or one of $(|\vec{p_4}|, |\vec{p_5}|)$) in terms of the other parameters. 
For example, the solution of $E_5$ can be expressed as
\begin{align}
E_{5, \pm}^0 = &\frac{1}{2 \left[ (2 E - E_4)^2 - (E_4^2 - m_f^2) f_{\theta}^2 \right]} \nonumber \\[0.3ex] 
&\hspace{0.5em} \times \biggl[ (2 E - E_4) (4 E^2 - 4 E E_4 + 2 m_f^2 - m_{\phi}^2) \nonumber \\[0.3ex]
&\hspace{2.4em} \pm \sqrt{(E_4^2 - m_f^2) F_5 (E, E_4, f_{\theta}^2) } \biggl] \, ,
\label{eq:E5sol} 
\end{align}
where we have defined the following function:
\begin{align}
F_5 (E, E_4, X) \equiv &\Bigl[ (4 E^2 - 4 E E_4 + 2 m_f^2 - m_{\phi}^2)^2 \nonumber \\[0.3ex]
&\hspace{1.0em} - 4 (2 E - E_4)^2 m_f^2 \Bigr] X \nonumber \\[0.5ex]
&\hspace{1.0em} + 4 m_f^2 (E_4^2 - m_f^2) X^2 \, .
\end{align}
Note that the relation obtained from Eqs.~\eqref{eq:p4p5comexplicit}, \eqref{eq:p4p5com} is symmetric under $E_4 \leftrightarrow E_5$ exchange. 
We emphasize that Eq.~\eqref{eq:E5sol} is originally from the $p_3^2 = m_{\phi}^2$ relation, and hence, we can use this solution for the integral of $E_5$ to obtain the cross section.

\subsection{Phase space integration}
\label{sec:phasespace}

To calculate the cross section for the process, we need to perform the phase space integration. 
In general, there are 9 parameters before applying the four-momentum conservation, and hence, we need to perform 9 integration for the three-body decay process:
\begin{align}
\nonumber
d \Phi_3 (p_1 + p_2; p_3, p_4, p_5) =& \left( \prod_{i = 3}^5 \frac{d^3 p_i}{(2 \pi)^3 2 E_i} \right) \\ & \times \delta^4\left( p_1 + p_2 - p_3 - p_4 - p_5 \right) \, .
\label{eq:Phi3}
\end{align}
The delta functions reduce the number of integrals to 5, and in our calculation, we choose independent parameters as
\begin{align}
E_4 \, , ~ \thf \, , ~ \thfb \, , ~ \vpf \, , ~ \vpfb \, .
\end{align}
Note that $E_5$ is parameterized by Eq.~\eqref{eq:E5sol}, and $p_1$ and $p_2$ are written by input parameters, the beam energy $E$ and electron mass $m_e$. 
To deal with the delta functions appropriately, we can use the following relation:
\begin{align}
\int \! \frac{d^3 \vec{p_i}}{2 E_i} = \int \! d^4 p_i \theta (E_i) \delta \left( p_i^2 - m_i^2 \right) \, .
\end{align}
Then, integration over $p_3$ can be done by the delta functions in Eq.~\eqref{eq:Phi3}, which gives
\begin{align}
d \Phi_3 (p_1 + p_2; p_3, p_4, p_5) &= \frac{d^3 p_4 d^3 p_5}{(2 \pi)^9 4 E_4 E_5} \theta (2 E - E_4 - E_5) \nonumber \\ & \times \delta \left( (p_1 + p_2 - p_4 - p_5)^2 - m_{\phi}^2 \right) \, .
\end{align}
$d^3 p_4 d^3 p_5$ can be replaced by
\begin{align}
    d^3 p_4 d^3 p_5 &= \big|\det{\cal J}\big| |\vec{p_4}| d \thf d \vpf d |\vec{p_5}| d \thfb d \vpfb \nonumber \\[0.6ex]
&= |\vec{p_4}|^2 |\vec{p_5}|^2 s_{\thf} s_{\thfb} d |\vec{p_4}| d \thf d \vpf d |\vec{p_5}| d \thfb d \vpfb \ ,
\end{align}
with the Jacobian matrix ${\cal J} \equiv \frac{\del (p_{4x},p_{4y},p_{4z},p_{5x},p_{5y},p_{5z})}{\del (|\vec{p_4}|,|\vec{p_5}|,\theta_f, \vpf, \thfb, \vpfb)}$.

The integration over $\vpf$ gives $2 \pi$, because all inner dots and $E_5$ solution in Eq.~\eqref{eq:E5sol} are independent of $\vpf$. 
$d |\vec{p_i}|$ can be replaced by $d E_i$ as
\begin{align}
d |\vec{p_i}| = \frac{E_i d E_i}{\sqrt{E_i^2 - m_i^2}} = \frac{E_i d E_i}{|\vec{p_i}|} \, .
\end{align}
The delta function is
\begin{align}
0 &= (p_1 + p_2 - p_4 - p_5)^2 - m_{\phi}^2 \nonumber \\[0.5ex]
&= 4 E^2 - 4 E (E_4 + E_5) + 2 E_4 E_5 + 2 m_f^2 \nonumber \\ &\hspace{1.2em} - 2 \sqrt{E_4^2 - m_f^2} \sqrt{E_5^2 - m_f^2} f_{\theta} - m_{\phi}^2 \nonumber \\
&\equiv f_{\delta}(E_5) \, ,
\end{align}
and hence, it is easy to perform the $E_5$ integration by using the solution in Eq.~\eqref{eq:E5sol}. 
For this calculation, we need to take into account the Jacobian appropriately, and in this case,
\begin{align}
& \delta \left( (p_1 + p_2 - p_4 - p_5)^2 - m_{\phi}^2 \right)
= \frac{\delta ( E_5 - E_5^0 )}{\left| f_{\delta}'(E_5^0) \right|}
\nonumber \\[0.5ex]
&= \left| 2 E_4 - 4 E - \frac{2 E_5^0 \sqrt{E_4^2 - m_f^2} f_{\theta}}{\sqrt{E_5^0 - m_f^2}} \right|^{-1} \delta ( E_5 - E_5^0 ) \, ,
\end{align}
with the solution $E_5^0$. 
Finally, we have
\begin{align}
d \Phi_3 (p_1 + p_2; p_3, p_4, p_5) &= \frac{\sqrt{E_4^2 - m_f^2} \sqrt{(E_5^0)^2 - m_f^2}}{(2 \pi)^8 4 \left| f_{\delta}'(E_5^0) \right|} \nonumber \\ & \times d E_4 d c_{\thf} d c_{\thfb} d \vpfb \, .
\end{align}
Hence, the result of the differential cross section for the process is
\begin{align}
d \sigma &= \frac{(2 \pi)^4}{4 E^2 |\vec{v_1} - \vec{v_2}|} \overline{\left| \cal{M} \right|^2} d \Phi_3 (p_1 + p_2; p_3, p_4, p_5) \nonumber \\[0.5ex]
&= \frac{\sqrt{E_4^2 - m_f^2} \sqrt{(E_5^0)^2 - m_f^2}}{32 E^2 (2 \pi)^4 \left| f_{\delta}'(E_5^0) \right|} \overline{\left| \cal{M} \right|^2} d E_4 d c_{\thf} d c_{\thfb} d \vpfb \, ,
\label{eq:dsigma}
\end{align}
where $\overline{\left| \cal{M} \right|^2}$ is corresponding averaged squared amplitude (with appropriate replacement for $E_5 \to E_5^0$), $|\vec{v_1} - \vec{v_2}|$ is the relative velocity of the beam, which is $2$ for COM frame.

\bibliography{bib}

\end{document}